\documentclass[12pt]{article}
\usepackage{epsfig,axodraw,amsbsy,latexsym}
\def\aka{{\it a.k.a.}\ }
\def\ie{{\it i.e.}\ }
\def\eg{{\it e.g.}\ }
\def\etc{{\it etc.}\ }

\def\cf{{\it cf.}\ }
\def\viz{{\it viz.}\ }
\def\be{\begin{equation}}
\def\ee{\end{equation}}
\def\bea{\begin{eqnarray}}
\def\eea{\end{eqnarray}}

\def\phi{{\varphi}}
\def\etab{{\bar\eta}}
\def\altpsi{\vartheta}
\def\f{{\rm f}}
\def\g{{\rm g}}
\def\gt{\tilde{g}}
\def\epsilont{\tilde{\epsilon}}
\def\del{\delta}
\def\DG{{\cal D}_\Gamma}
\def\sig3{\sigma_3}
\def\Lam{\Lambda}

\def\SUNN{$SU(N|N)$}
\def\bSUNN{$\boldsymbol{SU(N|N)}$}
\def\SUNM{$SU(N|M)$}
\def\X{{\cal X}}
\def\Xdot{\dot{\cal X}}
\def\vac{|0\rangle}

\def\A{{\cal A}}
\def\S{{\cal S}}
\def\Amu{{{\cal A}_{\mu}}}
\def\Bbar{\bar{B}}
\def\C{{\cal C}}
\def\F{{\cal F}}
\def\H{{\cal H}}
\def\J{{\cal J}}
\def\K{{\cal K}}
\def\L{{\cal L}}
\def\P{{\cal P}}
\def\Q{{\cal Q}}

\def\zetab{\bar{\zeta}}
\def\Dbar{\bar{D}}
\def\gap{\hspace{0.05in}}
\def\dint{\int\!d^D\!x}
\def\c{c^{-1}}
\def\ctil{\tilde{c}^{-1}}
\def\ct{\tilde{c}}
\def\chat{\hat{c}^{-1}}
\def\ch{\hat{c}}
\def\rtil{\tilde{r}}
\def\rt{\rtil}
\def\rtilo{\tilde{r}_0}
\def\rhat{\hat{r}}
\def\rh{\rhat}
\def\rhato{\hat{r}_0}
\def\tr{\mathrm{tr}}
\def\str{\mathrm{str}}
\def\wick#1#2{\hspace{-#1mm}\raisebox{-2ex}{\rule{0.02mm}{2mm}\rule{#2mm}
{0.02mm}\rule{0.02mm}{2mm}}\hspace{#1mm}\hspace{-#2mm}}
\def\one{\hbox{1\kern-.8mm l}}

\def\ds{\displaystyle}

\def\eq#1{(\ref{#1})}
\def\sec#1{sec.\ \ref{#1}}
\def\dga{${\cal D}_{\Gamma}$}
\def\dgaone{{\cal D}_{\Gamma}^{\mbox{{\scriptsize 1-loop}}}}

\begin{document}
\begin{titlepage}
\begin{flushright}
SHEP 01-15 \\
hep-th/0106258
\end{flushright}
\vspace{.4in}

\begin{center}
\renewcommand{\thefootnote}{\fnsymbol{footnote}}
{\Large{\bf Gauge invariant regularisation via \bSUNN}}
\bigskip \\ Stefano Arnone, Yuri A. Kubyshin,\footnote{Permanent address:
Institute of Nuclear Physics, Moscow State University, Moscow 119899, 
Russia} Tim R. Morris,  and John F. Tighe\\
{\it
Department of Physics, University of Southampton, Highfield\\
Southampton SO17 1BJ, UK\\
E-mails}: { \tt sa@hep.phys.soton.ac.uk, kubyshin@hep.phys.soton.ac.uk,
T.R.Morris@soton.ac.uk, jft@hep.phys.soton.ac.uk}

\vspace{.5in}
{\bf Abstract}\bigskip \end{center} \setcounter{page}{0}
We construct a gauge invariant regularisation scheme for pure $SU(N)$
Yang-Mills theory in 
dimension four or less (for $N=\infty$ in
all dimensions), with a physical cutoff scale $\Lambda$, by using
covariant higher derivatives and spontaneously broken \SUNN\ 
supergauge invariance.  Providing their powers are within certain
ranges, the covariant higher derivatives cure the superficial
divergence of all but a set of one-loop graphs. The finiteness of these
latter graphs is ensured by properties of the supergroup and gauge
invariance.  In the limit $\Lambda\to\infty$, all the regulator fields
decouple and unitarity is recovered in the renormalized pure $SU(N)$
Yang-Mills theory.  By demonstrating these properties, we prove that
the regularisation works to all orders in perturbation theory.
\end{titlepage}

\section{Introduction} 
\label{Introduction}

\addtocounter{footnote}{-1}
Nowadays the exact renormalization group approach (ERG) 
\cite{WH}, \cite{Polch}, which is essentially a continuous version 
of the Wilson renormalization group, is widely 
recognized as a powerful tool for {\it non-perturbative} calculations 
in quantum field theory (see refs.~\cite{catchall}-\cite{BTW} for 
reviews). Central objects within this approach are the effective action 
$S_{\Lambda}[\phi]$, where $\phi$ is a generic notation for the fields 
in the theory, and the ERG equation. $S_{\Lambda}[\phi]$ describes 
the physics of the theory in terms of fields and parameters (masses and 
coupling constants) relevant to the energy scale $\Lambda$. The ERG 
equation determines the change of the effective action with the 
change of the scale thus allowing, in principle,  physics at 
different scales to be related. 
One of the ambitious objectives of the approach is to trace the 
emergence of the low energy limit of the theory of strong interactions 
and to describe the chiral phase transition starting from the QCD 
action at high $\Lambda$.    

Though technically the ERG approach sometimes may seem to be more 
complicated than perturbation theory, the main idea behind it is 
quite simple. $S_{\Lambda}[\phi]$ effectively includes only field 
modes with momenta $|p| < \Lambda$. This is achieved by modifying 
the propagator as 
\be
     \frac{1}{p^{2}} \rightarrow 
     \frac{c \left( p^{2}/\Lambda^{2} \right)}{p^{2}}, 
               \label{p-cp}
\ee
where $c(p^{2}/\Lambda^{2})$ is a regulating function, or simply a 
regulator. The main requirements on this function are: 
(1) $c(z) \rightarrow 0$ 
as $z \rightarrow +\infty$ fast enough; (2) $c(0) = 1$. 
The precise form of the cutoff function is to a large extent the
choice of the practitioner, made for example to improve the convergence
of non-perturbative approximations \cite{Tigger}. In many 
calculations the sharp cutoff defined by $c(z)=1$ for $0 \leq z < 1$ and 
$c(z)=0$ for $z \geq 1$ was used \cite{WH}, \cite{HH}. However, smooth 
cutoffs were shown to have many advantages and to be more preferable 
for practical calculations \cite{Polch}, \cite{Tigger}, \cite{BaTh}. 
In this case higher momentum modes are not cut off exactly but 
suppressed. For many studies, setting 
$1/c(z)$ to be a polynomial function in $z$ serves the purpose
\cite{deriv}, \cite{erg2}. 
The ERG equation is an integro-differential equation containing 
variational derivatives of the effective action which is of the 
following generic form\footnote{There is ambiguity in the form 
of the ERG equation, due to the reparametrization invariance
\cite{LaMo00}.}: 
\be
- \Lambda \frac{d S_{\Lambda} [\phi]}{d \Lambda} = 
{\cal F} \left(S_{\Lambda} [\phi], \frac{\delta S_{\Lambda}}{\delta \phi_{p}}, 
\frac{\delta^{2} S_{\Lambda}}{\delta \phi_{p} \delta \phi_{q}}; 
c 
\right), 
         \label{ERGeq}
\ee
where ${\cal F}$ is some functional and $\phi_{p}$ stands for the 
field in the momentum representation. Once the regulator is introduced, 
eq. (\ref{ERGeq}) is well defined in the {\it non-perturbative} 
sense. Furthermore, the complete physical 
information of the theory may be extracted from the $\Lambda\to0$
limit of the effective action (\eg via its relation \cite{trm1}
to the Legendre effective action \cite{leg}). In this way,
the ERG equations have proven to be very useful and efficient in 
performing {\it non-perturbative}, albeit approximate in concrete 
studies, calculations of many effective physical properties 
in scalar and fermionic theories. The
non-perturbative treatment of gauge theories within the ERG approach 
was afflicted however with a long standing problem, the main 
difficulty being to 
construct a regularised effective action with a regulator suppressing 
momentum modes above the scale $\Lambda$ and, at the same time, 
maintaining gauge invariance. A version of the ERG equation satisfying 
this criterion to one loop,
was proposed in refs.~\cite{manerg}--\cite{erg2},
for $SU(N)$ gauge theory in the large $N$ limit. 
However the problem of constructing a scale dependent cutoff (or 
regularisation) suitable for gauge invariant ERG and without these
limitations, remained unsolved. 

The regularisation in a gauge invariant ERG formalism must meet the 
following requirements: 
\begin{enumerate} 
\item be gauge invariant; 
\item perform the (effective) suppression of higher momentum 
field modes, thus implementing one of the basic concepts of the ERG; 
\item allow a non-perturbative treatment. 
\end{enumerate}
The last requirement means that the regularisation does not rely 
on the diagram expansion, but does its job within the ERG equation as 
such, and is therefore of the utmost importance for the ERG approach. 

In the scalar case a simple modification of the propagator
in the effective action, as given by 
(\ref{p-cp}), is sufficient to regularise the 
theory \cite{Polch}, \cite{BaTh}. Namely,  ultraviolet divergences 
appear neither in non-perturbative studies, nor in perturbative 
computations 
(of course, perturbative calculations can always be carried out 
using the ERG equation). In the gauge case the situation is more 
complicated. A straightforward modification of the pure gauge 
(i.e. Yang-Mills) action to incorporate the regulator even in the gauge 
covariant form $c(-\nabla^{2}/\Lambda^{2})$ (see details below) turns 
out to be insufficient. Namely, after such a modification 
certain ultraviolet divergences remain in the theory. 
A solution to this problem, proposed in some 
earlier papers \cite{erg1}, \cite{tver}, \cite{rome}, 
is to extend the theory by incorporating some 
other (non-physical) fields. Of course, when the regularisation 
is removed, i.e. in the limit $\Lambda \rightarrow \infty$, the 
extended theory must reduce to the initial Yang-Mills theory. 

In this paper we construct a Poincar\'e invariant and gauge invariant 
ultraviolet cutoff which works for pure $SU(N)$ Yang-Mills theory 
in 
dimension $D\le4$. In the large $N$ limit the regularisation
works for all dimensions. 
There are a number of reasons in fact why such a regularisation
may prove useful. As already explained, our 
primary motivation is that this regularisation may be 
incorporated into an elegant and potentially powerful manifestly gauge 
invariant exact renormalization group 
\cite{manerg}--\cite{romerg}, \cite{us}. 
As stated above,
the key features required of such a regulator are that it is gauge
invariant, introduces a mass scale, and that it makes sense 
non-perturbatively.
Although we only prove here that the regularisation works to all orders
in perturbation theory, it is natural to expect that it
is valid non-perturbatively since it is based on a physical cutoff, \ie
a real ultraviolet cutoff scale $\Lambda$ 
\cite{tver,rome}.\footnote{as opposed to \eg
analytic behaviour of perturbative amplitudes as in  dimensional 
regularisation}  
Our regularisation should be compared in these respects to the only two 
non-perturbative regularisations known so far: the lattice
regularisation, which of course is not formulated in the continuum thus
breaking Poincar\'e invariance, and Slavnov's higher derivative scheme
\cite{pvo,pvs,pva}. We must stress that our purpose is not to construct 
a new regularisation for perturbative computations in Yang-Mills theories 
(though, of course, it can be checked that our regularisation works 
in perturbation theory as well and gives correct results). 
There are many powerful and effective techniques for diagram 
calculations. Our goal is to have a regularisation which 
allows {\it non-perturbative} treatment of gauge theories 
within the ERG approach.

Nevertheless, ideologically our regularisation is quite close to 
Slavnov's covariant higher derivative scheme elaborated in 
refs.~\cite{pvo,pvs,pva}. We review this and other earlier work, and 
provide all the details, below. First we sketch the basic idea, which is 
very simple. (See also refs. \cite{tver,rome}.)
Covariant higher derivative regularisation
of $SU(N)$ Yang-Mills does not
work on its own because one-loop divergences slip through \cite{cov}.
We can cure this problem by working instead with
$SU(N|N)$ Yang-Mills which has sufficiently improved ultraviolet 
properties. 


$SU(N|N)$ has as subgroup $SU(N)\times SU(N)$. 
Correspondingly the \SUNN \ super-gauge field, $\A$, contains two 
gauge fields transforming separately under each $SU(N)$, 
a normal one, $A^1$, which will be the physical gauge field, 
and a copy, $A^2$,
with wrong sign action. $\A$ also contains
a complex fermionic gauge field $B$ that transforms as a
fundamental under one $SU(N)$ and complex conjugate fundamental 
under the other.\footnote{There can be also a
decoupled $U(1)$ gauge field $\A^0$, depending on how one represents
$SU(N|N)$. We cover such subtleties later.}
The remaining potential
divergences are cancelled via the (linear representation of)
supersymmetry in the fibres\footnote{as opposed to spacetime supersymmetry
formed by 
non-linear representation from the 
super-Poincar\'e over super-Lorentz coset space. This in turn
leads to supermatrix valued fields as opposed to superfields valued on a 
superspace.} of the unbroken \SUNN :
in these cases for every purely bosonic loop, there is also a  
purely fermionic one which is an exact copy but which
enters with the opposite sign. For mixed bosonic/fermionic loops the
wrong sign $A^2$ gauge field propagators ensure cancellation.  


Of course neither $B$ nor $A^2$
is physically meaningful. By introducing a superscalar
Higgs' field which spontaneously breaks
the supersymmetry in the fermionic directions, we can give
arbitrarily high masses to the fermionic field. $B$ then behaves
exactly like a massive Pauli-Villars field and it is only through
this that the two $SU(N)$ gauge fields can interact. Since cancellation
will still take place but now only at large loop momenta, 
in effect a new physical cutoff has been introduced that suppresses 
high momentum modes. Initially we verified these mechanisms by working
explicitly in components, however in this paper
the work is presented using the full superfields, so this 
Fermi-Bose interpretation will remain hidden
just below the surface. The cancellations will arise through a supergroup
theoretical analogue: the `supertrace mechanism'. In particular, quantum 
corrections that yield $\tr\one=N$ in $SU(N)$ Yang-Mills now yield 
$\str\one=0$.

Note that apart from $A^1$, only 
the unphysical $A^2$ field remains massless. We need
to verify that no effective interaction is left between $A^1$ and
$A^2$ as the symmetry breaking scale $\Lambda$ 
is sent to infinity, so that we
can simply ignore the non-unitary $A^2$ sector. But this is
guaranteed by the Appelquist-Carazzone theorem providing the theory
is renormalizable \cite{apple} (which is where the restriction to dimensions
$D\le4$ comes in) since the lowest dimension gauge invariant
effective interaction involves the square of the two field strengths:
\be
\label{12eff}
\tr(F^1)^2\,\tr(F^2)^2
\ee
and is thus irrelevant, vanishing as an inverse power of $\Lambda$.

The necessary double trace makes all such terms subleading in large
$N$, thus decoupling takes place in any dimension in the large $N$ limit
\cite{thooftln}. Furthermore, in this limit the supertrace mechanism 
ensures that the theory is finite in all dimensions (otherwise
by the same token \eq{12eff},
or rather their supergroup version, is the lowest dimension term
not suppressed by the supertrace mechanism at one-loop, 
resulting in a divergence in $D\ge8$ dimensions). These are the reasons
why the $N=\infty$ limit of $SU(N)$ Yang-Mills is regulated by our
theory in all dimensions $D$.

Our regularisation scheme was initially
inspired by earlier work of Slavnov and others, on covariant higher 
derivatives. As already remarked, it is a well established problem that
higher covariant derivatives 
fail to cure ultra-violet divergences at one loop \cite{cov}.
The one-loop divergences can be regularised by also introducing
gauge invariant Pauli-Villars (PV) fields \cite{pvo}, the action being bilinear 
in these fields so that they provide, on integrating out, the missing one
loop counterterms (plus other finite contributions). But further 
one-loop divergences then typically arise when the PV
fields are external. 
Whilst these might be ignored on the grounds that the Pauli-Villars quanta
are not to be regarded as external physical particles, the divergences
reappear in internal subdiagrams as overlapping divergences at
higher loops \cite{pvp}. 
Further controversy was caused by the discovery of unphysical
contributions arising from, as it turned out, an unnecessary restriction
to ``covariant Landau gauge'' \cite{pvc,pvs,pva}. 

These difficulties have been resolved by adding more PV
fields and by judicious choices for their actions \cite{pvs,pva}. Even so,
there are problems: the solution is unwieldy, and inappropriate for 
incorporation in the exact renormalization group framework since it is not 
possible to preserve the property that the PV fields appear only bilinearly
at the level of the effective action \cite{manerg,erg2}. Instead, we needed and 
here furnish, a framework in which the gauge fields and gauge invariant
PV fields are treated on the same footing. From the start the regularisation 
applies to all fields simultaneously, and thus the above problem of 
``overlapping divergences'' never arises. It also means that there is 
simply no analogue of the covariant Landau gauge controversy.  

An earlier $N=\infty$ version of the present framework
was constructed by adding such PV fields by hand but insisting that the 
resulting regularisation respected the  exact renormalization group flow 
\cite{catchall}, which in particular meant that higher order interactions
for the PV fields had to be added \cite{manerg}. The result regularised only
the one-loop diagrams without external PV fields \cite{erg2}, 
\ie still suffered the above problem of ``overlapping divergences''.
Nevertheless, it was realised that this version
could be understood at a deeper level as arising from 
a spontaneously broken \SUNN \ 
gauge theory, albeit in a form of unitary gauge and with some small
differences, and this led to the suggestion that  
higher derivative regularisation based exactly on spontaneously broken \SUNN\ 
may work to all loop orders and also at finite $N$ \cite{erg2}. 

In implementing these ideas we uncovered
numerous novel features. In sec. \ref{SUNN}, we meet one 
of the causes. \SUNN\ is reducible but indecomposable:
it has a bosonic $U(1)$ subalgebra in the $\one$ direction,
which thus commutes with everything,
but which cannot be discarded because
it is itself generated by fermionic elements of the algebra. 
As we see in \sec{Alternatives}, for the gauge
theory this means that there is a $U(1)$ gauge field $\A^0$ with no
kinetic term and
which interacts with nothing, but is nevertheless necessary to ensure
gauge invariance! We show in \sec{Alternatives}
how to construct two equivalent representations
in one of which the $\A^0$ is `projected out'. 
We also note that its decoupling is protected by a simple shift symmetry. 
In fact, we step back
to consider also $U(N|N)$ which does not factor, even locally, into
$SU(N|N)\times U(1)$. We find that a gauge theory built on
$U(N|N)$ automatically contracts itself to $SU(N|N)$! We also carefully
consider the implications of these and similar peculiarities for the superscalar
sector. We finish this section discussing 
another novelty: the ghost degrees of
freedom do not (anti)commute sensibly with the supergroup directions
leading to a breakdown of global $SU(N|N)$ invariance and failure of
at least na\"\i ve implementations of BRST.
We furnish an elegant solution by going beyond the simple distinction of 
fermion or boson and introducing two separate gradings for ghost and 
supergroup degrees of freedom. 

In \sec{Spontaneously} we present our higher derivative regularised
theory, its form in the appropriate 't Hooft gauge and the resulting ghost
action. In \sec{Counting}, we ignore the special cancellations provided by 
the supergroup and find necessary and sufficient 
conditions on the powers of the higher derivatives (in fact ranks of
polynomials of these) to regularise as many diagrams as 
possible. Whilst furnishing sufficient conditions and isolating a set
of `One-loop Remainder Diagrams' which still need further regularisation,
follows quickly from standard power counting methods, finding the minimal
set of sufficient conditions requires more cunning. In \sec{Supertrace}
we prove that two and three-point One-loop
Remainder Diagrams are regularised by the supertrace mechanism. 
We also show that all but the 
unbroken parts of the One-loop Remainder Diagrams are superficially finite 
by covariant higher derivatives alone. We finish by noting that in the large 
$N$ limit the symmetric phase has no quantum corrections at all, and thus
in this limit the theory is finite in all dimensions $D$.
In \sec{Ward} we turn to the remaining One-loop Remainder Diagrams 
at finite $N$, which are ultraviolet finite only after gauge invariance is 
taken into account, by a combination of the higher derivatives and the 
supertrace mechanism, in all dimensions $D<8$. Actually since
we are dealing with the gauge fixed theory these arguments need to be
phrased in terms of Ward identities and
BRST invariance which we develop here.  
With this structure in place we complete
the proof of finiteness to all orders in perturbation theory
of covariant higher derivative regularised 
spontaneously broken $SU(N|N)$ in all dimensions $D<8$. 


In order for this to act as a regulator for $SU(N)$ Yang-Mills
we are left to show that the low energy sector is given by $SU(N)$
Yang-Mills. There is a case to answer because the wrong-sign $A^2$
field remains massless. In \sec{Unitarity} we first confirm that the wrong sign
leads to negative probabilities and a non-unitary S-matrix for this sector
and then prove that the two sectors decouple in $D\le4$ dimensions,
or at $N=\infty$ in all dimensions, as already sketched above.

Sec. \ref{Preregularisation} is devoted to the subtle issue of preregularisation.
Any PV regularisation achieves finite results
by the addition of separately divergent quantities. Thus the resulting integrals
are only conditionally convergent, and require some precise prescription if they
are to be unambiguously defined. 
One convenient possibility is to preregularise with
dimensional regularisation, which as we stress, makes sense as a preregulator
even non-perturbatively. However, in the $N=\infty$ case and/or when the
dimension $D<4$, {\sl no} preregularisation is necessary: the structure of 
$SU(N|N)$ group theory is sufficient to organise these integrals 
into absolutely convergent pieces which can then be evaluated unambiguously.
We would like to stress that the preregularisation is essentially needed 
only to obtain a precise definition of the One-loop Remainder Diagrams and the 
Ward identities. In practical calculations within perturbation 
theory, all potentially 
divergent terms cancel out automatically inside the integrands due to the 
supertrace mechanism. We expect that the same mechanism makes the
regularised theory finite in non-perturbative calculations as well.

Finally in \sec{Conclusions} we summarise and draw our conclusions.

\medskip
The controversy \cite{pvr,pvc} caused by Slavnov's initial choice of covariant 
Landau gauge \cite{pvo}, still provokes disquiet amongst some researchers
in the field, despite the elegant explanation \cite{pva} and solution 
\cite{pvs,pva} already published. Although not directly relevant to
the present research, we revisit it in appendix \ref{Landau}, to provide an even
clearer proof that the covariant Landau gauge was responsible for the 
appearance of unphysical contributions, by hiding a massless mode,
and to emphasise that it was this
restriction that caused the problem, not some generic difficulty with the
gauge invariant PV idea. This also allows us to demonstrate clearly that our
regularisation method has no analogous problem. 

\section {\bSUNN}
\label{SUNN}

We start with an elementary exposition of the $SU(N|N)$ superalgebra 
\cite{bars} and its invariants, covering the notation and key formulae
needed later on.
The defining representation of the graded Lie algebra of $U(N|M)$
is constructed from commutators of
Hermitian $(N+M)\times(N+M)$ matrices of the form:
\bea 
{\cal H} = \left(\begin{array}{cc} H_N & \theta \\
\theta^\dagger & H_M\end{array}\right),                 
\eea
where $H_N \, (H_M)$ is an $N\times N \, (M \times M)$ Hermitian matrix with complex
bosonic elements and  $\theta$ is an
$N\times M$ matrix composed of  complex Grassmann numbers. $\H$ is thus a 
Hermitian supermatrix. The supertrace replaces the trace
as the natural invariant for supermatrices:
\be 
\str({\cal H})= \tr(\sig3{\cal H})
= \tr(H_N) - \tr(H_M), 
\label{supertrace1}
\ee
where 
\be
\label{sigma3}
\sig3= \left(\!\begin{array} {cc} \one_N & 0 \\ 0 & -\one_M
\end{array} \!\!\right),\ee
with $\one_N$ ($\one_M$) being the $N\times N$ ($M\times M$)
identity matrix. This is because only the supertrace is cyclically
invariant (compensating the signs picked up by anticommuting Grassmann
components):
\be
\label{cycle}
\str\, XY=\str\, YX
\ee
(where $X$ and $Y$ are two general supermatrices),
ensuring the supertrace of a commutator vanishes, and thus in turn 
ensuring invariance under adjoint action of the group.

We will define the generators of the group
to be Hermitian matrices with only complex number entries, the Grassmann
character being carried as appropriate by the coefficients (the 
superangles). In 
terms of the generators then, we obtain a superalgebra with commutators
or anticommutators as appropriate.

To be elements of
the algebra of \SUNM \  we require that ${\cal H}$ be supertraceless, \ie
$\str\,\H=0$.
With the traceless parts of $H_N$ and $H_M$ corresponding to $SU(N)$ and 
$SU(M)$ respectively and the orthogonal traceful part giving rise to a 
$U(1)$, we see that the bosonic sector of the \SUNN \  algebra  forms a 
$SU(N)\times SU(M)\times U(1)$ subalgebra.

We now specialise to $M=N$.  In this case the algebra is reducible because
the bosonic $U(1)$ subalgebra is generated by the unit matrix $\one_{2N}$
which thus 
commutes with all the other generators, forming an Abelian ideal
(invariant subspace). In contrast to compact bosonic Lie algebras however,
we cannot then decompose $SU(N|N)$ into a direct product of smaller 
algebras, because $\one_{2N}$ is itself generated by fermionic elements of 
the superalgebra, for example 
\bea
\label{why1}
\left\{\sigma_1,\sigma_1\right\} &=& 2\one_{2N},\\
{\rm where}\qquad\sigma_1 &=& \pmatrix{0 & \one_N\cr \one_N & 0}\label{sig1}.
\eea
Ref. \cite{bars} excludes $\one_{2N}$
by redefining the Lie bracket in this representation. 
As we explain in sec. \ref{Alternatives}, it will turn
out that in constructing our action, we cannot exactly exclude $\one_{2N}$
in this way, and thus our definition of \SUNN\
is different from that of ref. \cite{bars}. 
We do note though that the unit matrix has a special role
to play and for this reason we separate it from the other generators.

We define the generators, $S_\alpha \equiv \{\one , T_A \}$,
where the $T_A$ are complex block diagonal and block off-diagonal 
Hermitian traceless matrices. $A$ runs
over the $2(N^2-1)$ bosonic (\aka block diagonal) generators
and $2N^2$ fermionic (\aka block off-diagonal) generators and $\alpha
\equiv\{0,A\}$. An element of the \SUNN \ algebra is then
\bea
\label{gensum}
\H & = &  \H^{\alpha} \, S_\alpha =\H^0\one+\H^A T_A,\nonumber \\
({\cal H})^i_{\gap j} & = & \H^{0}\,\del^i_{\gap j} + \H^A\,(T_A)^i_{\gap j}.
\eea
enabling us to identify the Killing super-metric
\be
h_{\alpha\beta} = 2\,\str(S_{\alpha}S_{\beta}).
\ee
This is symmetric when $\alpha$ and $\beta$ are both bosonic and 
antisymmetric when both are fermionic, \ie
\be\label{metricasym}
h_{\alpha\beta} = h_{\beta\alpha}(-1)^{\f(\alpha)\f(\beta)},
\ee
where $\f(\alpha)$ is $0$ when $\alpha$ is bosonic and $1$ when it is
fermionic. The generators are normalised such that
\bea
\label{killing}
h_{\alpha\beta} =
\left( 
\begin{tabular}{c|ccc|ccc|ccccc}
$0$& &&& &&& &&&& \\ \hline
& $1$&&& &&& &&&& \\ 
& &$1$&& &&& &&&& \\ 
& &&$\ddots$& &&& &&&& \\ \hline
& &&& $-1$&&& &&&& \\
& &&& &$-1$&& &&&& \\ 
& &&& &&$\ddots$& &&&& \\ \hline
& &&& &&& $0$& $i$& && \\ 
& &&& &&& $-i$& $0$& && \\ 
& &&& &&& && $0$& $i$&  \\ 
& &&& &&& && $-i$& $0$&  \\ 
& &&& &&& &&&& $\ddots$ \\
\end{tabular}
\right)\\
\underbrace{\hspace{0.8cm}}_{U(1)}\,
\underbrace{\hspace{2cm}}_{SU_1(N)}\,
\underbrace{\hspace{2.7cm}}_{SU_2(N)}\,
\underbrace{\hspace{4cm}}_{\mathrm{Fermionic}}
\hspace{0.2cm} \nonumber
\eea
Obviously, this has no inverse.  However, 
we can define the restriction of $h_{\alpha\beta}$ to the $T_A$ space:
\be
\label{gmet}
g_{AB} = 2\, \str(T_AT_B)=h_{AB},
\ee
with inverse determined by
\be
\label{igmet}
g_{AB}g^{BC}=g^{CB}g_{BA}=\del^C_{A}.
\ee
The metric can be used to raise and lower indices
\be
X_A :=g_{AB}X^B\quad\Longrightarrow\quad X^A=X_Bg^{AB} \neq X_Bg^{BA}.
\ee
Note that it is the second index of the metric that is summed over; from
(\ref{metricasym}) it is clear that the ordering of the indices of the
metric is important. Using \eq{metricasym} and the above relations one
may readily verify that for superangles $X^AY_A=Y^AX_A$ ($\neq X_AY^A$)
in agreement with \eq{cycle}. We can add the dual relations for raising
and lowering indices on generators:
\be
T^A :=T_B g^{BA}\qquad{\rm so}\qquad X^A T_A= X_A T^A
\ee
(with ordering of $X$ and $T$ of course unimportant). Finally,
the $T_A$ generators of \SUNN \   give rise to a completeness relation
\be\label{fierz0}
(T^A)^i_{\gap j} (T_A)^k_{\gap l} = {1\over 2}\del^i_{\gap
l}\, (\sig3)^k_{\gap j} - {1\over 4N}\left[\del^i_{\gap j}\,(\sig3)^k_{\gap
l} + (\sig3)^i_{\gap j}\,\del^k_{\gap l}\right],
\ee
which is most usefully cast contracted into arbitrary supermatrices $X$
and $Y$:
\bea
\label{sow}
\str(XT^A)\, \str(T_AY) &=&
{1\over2}\,\str\,XY-{1\over4N}\left(\tr X\str Y + \str X\tr Y\right),\\
\label{split}
\str(T^AXT_AY) &=&
{1\over2}\,\str X\,\str Y-{1\over4N}\tr\left(XY+YX\right).
\eea

\section{Alternatives and otherwise}
\label{Alternatives}

When treated as a gauge group, the
$SU(N|N)$ algebra has some unusual features which mean that a number
of steps in constructing the action have to be rethought from the beginning.
As we will see, the existence of fermionic directions present their own 
novelties particularly for the BRST algebra, but the main novelty is that,
as mentioned in sec. \ref{SUNN}, $SU(N|N)$ is an example of a superalgebra 
that is reducible but indecomposable.  


Actually, even the step of taking $SU(N|N)$ rather than $U(N|N)$ needs
to be carefully rethought: the reduction to $SU(N|N)$ is achieved by excluding
the space of generators with nonvanishing supertrace [spanned by adding
any representative \eg $\sigma_3$, as defined in \eq{sigma3}], but there 
is no corresponding ideal. (Note for example that $\sigma_3$ does not 
commute with the fermionic generators.) In other words, it is \emph{not} 
the case that the $U(N|N)$ group
is even locally isomorphic to $SU(N|N)\times U(1)$ !

Let us recall the reasons for treating separately parts of a 
reducible compact bosonic Lie group, for example $U(N)$. In this
case we know that we can locally decompose it into $SU(N)\times U(1)$, but 
the precise reason we treat the two subgroups separately in this context
is because the Lagrangian inevitably contains relevant 
pieces 
which are invariant separately under $SU(N)$ and $U(1)$. Since the 
$SU(N)$ piece and the $U(1)$ piece do not mix under the action of $U(N)$,
they will receive different divergent contributions and must therefore 
have their own couplings -- which renormalize separately.
Thus we see that reducibility matters for the renormalization of a 
Lagrangian and should be understood in terms of the possible invariants.
These issues are not directly relevant in the present case since we will be 
interested only in Lagrangians that result in a \emph{finite} theory.
Nevertheless, we will comment on the more general situation.


We start then by considering $U(N|N)$ and the pure gauge sector of the 
theory. Unless otherwise specified we will be working in $D$ Euclidean
dimensions. Extending \eq{gensum}, we write an element of the Lie 
superalgebra of $U(N|N)$ by extending the index to include the 
label ${}^\sigma$:
\be
\label{unn}
{\cal H}=\H^{\sigma}\sigma_3+\H^0\one+\H^A T_A.
\ee
Using $\tr\,T_A=0$,
the Killing supermetric \eq{killing} extends as follows:
\bea
h_{\sigma\sigma}\equiv2\str({\sigma_3}^2)&=&0,\nonumber\\
h_{A\sigma}=h_{\sigma A}\equiv2\str(\sigma_3 T_A)&=&0,\label{massacre}\\
h_{0\sigma}=h_{\sigma0}\equiv2\str(\sigma_3\one)&=&4N.\nonumber
\eea
Writing the covariant derivative as $\nabla_\mu=\partial_\mu-i\gt\A_\mu$,
where $\A_\mu$ is a member of the Lie superalgebra as in \eq{unn},
the field strength is 
\be
\label{fieldstrength}
\F_{\mu\nu}={i \over \gt}[\nabla_{\mu},\nabla_{\nu}].
\ee
The Lagrangian density is then
given by $\sim \str\, \F_{\mu\nu}^2$ (plus higher order interactions 
through extra commutators of $\nabla$, providing the covariant higher
derivative regularisation, \cf sec. \ref{Spontaneously}).

Recall from \sec{SUNN}, that we need the supertrace for its invariance
properties. Thus under gauge transformations
\be
\label{gaugetr}
\delta \A_\mu={1\over\gt}[\nabla_\mu,\omega]
\ee
(where $\omega$ is valued in the Lie superalgebra), we see that 
\be
\label{gaugecy}
\delta\, \str\, \F_{\mu\nu}^2 = -i\, \str\, [\F_{\mu\nu}^2,\omega],
\ee
with the r.h.s. (right hand side) 
vanishing after using cyclicity of the supertrace.

Actually, we can also construct the $U(1)$-like covariant derivative
\be
\label{u1covder}
\nabla^{(\sigma)}_\mu=\partial_\mu-i{\gt_\sigma\over2N}\str\A_\mu
=\partial_\mu-i\gt_\sigma\A^{\sigma}_\mu
\ee
and add the corresponding field strength squared. {\it A priori}, 
if we were dealing
with a divergent theory or were interested in finite renormalizations
within the $U(N|N)$ theory, we would not be allowed to exclude this term.
This much is similar to the case of $U(N)$ versus $SU(N)$, but note
that $\A^{\sigma}$ also appears in the kinetic term and interactions
in $\str\, \F_{\mu\nu}^2$, and contributes to the gauge transformations
of the other components via \eq{gaugetr}. The latter would mean that 
there is only one wavefunction renormalization, all components being
bound together via Ward identities. In fact we will see shortly that
the dynamics forces $\A^{\sigma}$ to disappear from the spectrum
so none of these curiosities need be pursued further here.

Since $\one$ commutes with everything, there are no $\A^0$ interactions. 
This is true even when we introduce adjoint matter fields.
Its only appearance is in the kinetic term as
\be
\sim -2N \A^0_\mu(\partial^2\delta_{\mu\nu}-\partial_\mu\partial_\nu)
\A^\sigma_\nu,
\ee
as follows from \eq{massacre}. (When covariant higher derivative
regularisation is included, an invertible polynomial $c^{-1}(-\partial^2/
\Lambda^2)$ is also inserted.)
Consequently $\A^0$ acts as a Lagrange
multiplier field, and integrating over it enforces the constraint that
$\A^{\sigma}$ is longitudinal \ie can be written
\be
\label{a3fate}
\A^{\sigma}_\mu=\partial_\mu\Omega^{\sigma}
\ee
for some $\Omega^{\sigma}$. 
(We will not consider the possibility that spacetime has non-trivial 
cohomology.) But under $U(N|N)$ gauge transformations $\A^{\sigma}$ changes 
like a $U(1)$ field
\be
\delta \A^\sigma_\mu={1\over \gt}\partial_\mu\omega^{\sigma},
\ee
receiving no contribution from $\A$ and 
the other generators because all graded commutators
in the superalgebra are supertraceless. Therefore \eq{a3fate} means
that  $\A^{\sigma}$ is constrained to be pure gauge.
We may as well pick the gauge corresponding to 
$\Omega^{\sigma}=0$ and thus get rid of 
$\A^{\sigma}_\mu$; the associated ghost
Lagrangian is free and decoupled and can be ignored. We see that the net
result is that even if we start with $U(N|N)$ Yang-Mills, it collapses to 
$SU(N|N)$ Yang-Mills! 

Without loss of generality we can \emph{start} with $SU(N|N)$ 
Yang-Mills, so $\A_\mu=\A^\alpha_\mu S_\alpha$ as in \eq{gensum}. Now
$\A^0$ does not appear in the Lagrangian at all!
Thus the $\A^0_\mu$ part of the partition function is a free functional 
integral, \ie without even a Gaussian weight,
contributing at most an infinite constant to the vacuum energy.
The lack of any sort of interaction involving $\A^0$ is guaranteed by
the appearance of a  new local symmetry $\delta\A^0_\mu(x)=\Lambda_\mu(x)$,
which we might as well call the ``no-$\A^0$'' symmetry since it is precisely
equivalent to the statement that $\A^0$ enters simply as a free functional
integral.
But we cannot simply exclude it because, as we saw from \eq{why1},
gauge transformations do appear in the $\one_{2N}$ direction, and
$\A^0_\mu$ must be there to absorb them! 

An alternative is to dispense with $\A^0$ by redefining
the Lie bracket to remove $\one$. We define a ``*bracket'' \cite{bars}
\be
[\ ,\ ]^*_{\pm}=[\ ,\ ]_{\pm}-{\one\over2N}\tr[\ ,\ ]_{\pm},
\ee
where $[\ ,\ ]_{\pm}$ is applied to the generators and is a commutator
or anticommutator as appropriate (so that passing to supermatrices as in
\eq{gensum}, they all become commutators). Note that the *bracket still 
satisfies the Jacobi identity. This follows because
\be
\label{jacobiminus}
[\H_1,[\H_2,\H_3]^*] = [\H_1,[\H_2,\H_3]]
\ee
(which in turn follows after noting that $\tr[\H_2,\H_3]$ is always bosonic)
and thus
\bea
\label{jacobi}
[\H_1,[\H_2,\H_3]^*]^* &=&[\H_1,[\H_2,\H_3]]^* \\
\nonumber &=& [\H_1,[\H_2,\H_3]]-{\one\over2N}\tr[\H_1,[\H_2,\H_3]],
\eea
for any elements $\H_i$ of the algebra. 
Since the *bracket is bilinear and antisymmetric and satisfies the 
Jacobi identities, it may equally well represent the Lie 
product. Using this, members of the Lie algebra may be written $\omega^AT_A$,
and thus $\A\equiv\A^AT_A$, and all Lie algebra commutators such as in 
\eq{fieldstrength} and \eq{gaugetr}, become *brackets. 

The *bracket simply sets to zero the structure constants that generated
$\one$, leaving all other structure constants alone, and because 
the Killing supermetric vanishes in the $\one$ directions,
none of the interactions change, as can be seen directly from 
the first of the two relations in \eq{jacobi} and 
\be
\label{jacobiplus}
\str\,\H_1[\H_2,\H_3]^*=\str\,\H_1[\H_2,\H_3],
\ee
which again holds for any elements $\H_i$ of the algebra.
Since the Lagrangian is actually unchanged by the introduction of the 
*bracket we see that physically the former ``free $\A^0$'' representation
and this latter *bracket representation are equivalent.

Of course the Lagrangian so far defined cannot represent an acceptable 
physical theory
not the least because the fermionic vector $B_\mu$ violates the spin
statistics theorem. But that is not our intention: instead we want to
give $B_\mu$ a mass of order the cutoff $\Lambda$, breaking the fermionic 
gauge invariance of the theory. Providing at energies 
much greater than $\Lambda$, the theory behaves like unbroken $SU(N|N)$, 
its finiteness properties will ensure that $B_\mu$ acts like a Pauli-Villars 
field cutting off energies in the unbroken $SU(N)\times SU(N)$ 
Yang-Mills\footnote{Only if we include $\A^0$,
there is also a free functional integral over this, but since we
have seen that it has no physical effect, we will not mention it further.}
above $\Lambda$. Fortunately we know how to do this, we must break the
theory \emph{spontaneously} in all and only the fermionic directions. The 
most general solution is to introduce a non-zero vacuum expectation value 
along a direction
\be
\label{break}
\sigma_3+\alpha\one
\ee
in the Lie superalgebra (where $\alpha$ can be any real number).

Thus we introduce a superscalar field $\C$ which, since it must live
in the Lie superalgebra containing $\sigma_3$, lies in the adjoint
of $U(N|N)$. Of course it is entirely consistent
for $\C$ nevertheless only to transform locally under $SU(N|N)$. It is the
fact that $U(N|N)\ne SU(N|N)\times U(1)$ that allows the theory to be
nevertheless nontrivial. Under gauge transformations \eq{gaugetr}, $\C$
will transform as
\be
\label{cgaugetr}
\delta\C = -i[\C,\omega].
\ee

Now, in the *bracket representation this commutator cannot be replaced by a
*bracket because the result would fail to be gauge invariant. To see this
consider the supertrace of an $n^{\rm th}$ order
monomial of adjoint representatives with $n>1$. The $\C$ kinetic term, 
which is necessary to give $B_\mu$ its mass, is an $n=2$ example. Another
example is $\str\, \C^n$ which we will use to construct a potential.
Clearly these are invariant under \eq{cgaugetr}.
But with $\delta\C = -i[\C,\omega]^*$, we obtain
\be
\delta\,\str\,\C^n={in\over2N}\str\,\C^{n-1}\ \tr[\C,\omega].
\ee
which is non-vanishing in general. (For $n=2$ it is non-vanishing
in general only if $\C$ contains $\sigma_3$, as indeed it does here.)
The underlying problem is that, while the *bracket is a perfectly fine
representative of the super-Lie product, we need it to be defined also
in the universal enveloping algebra (effectively here, general products of 
adjoint fields). Gauge invariance then requires the Leibnitz rule
\be
[\H_1,\H_2\H_3]=[\H_1,\H_2]\H_3+\H_2[\H_1,\H_3],
\ee
because 
it is this that implies $\delta\, \C^n=-i[\C^n,\omega]$. But the 
Leibnitz identity fails for the *bracket. 

We see that unlike the case for $\A$, we cannot exclude the $\one$ direction
from $\C$, which thus expands as
\be\label{Cexp}
\C=  \C^{\sigma}\sig3 + \C^{0}\one + \C^A\,T_A.
\ee
We can still dispense with $\A^0$ however as follows:
we use the *bracket for all pure gauge interactions as already described
above, but commutators are required when $\nabla$ acts on $\C$, \eg
in the superscalar's kinetic term 
\be
\label{cke}
\str\, [\nabla_\mu,\C]^2.
\ee
We cannot use a *bracket here because this time the
non-$\one$ interactions \emph{are} altered and the
result is not gauge invariant. This
follows from the breakdown of \eq{jacobiplus}
when $\H_1$ contains $\sigma_3$,
\ie is an element of $U(N|N)$ (in turn
the result of non-vanishing $h_{0\sigma}$).
To summarise: in our *bracket representation
$\A$ is a representative of $SU(N|N)$ without $\one$; $\C$ represents
$U(N|N)$ containing $\one$.
Under gauge transformations, $\A$ transforms with a *bracket in \eq{gaugetr},
but $\C$ transforms with a commutator as in \eq{cgaugetr}. The result is
consistent by relation \eq{jacobiminus} and the fact that $\C$ transforms
only into itself under \eq{cgaugetr}.
Trivially, the equality of the Lagrangian in the *bracket and free-$\A^0$
representations carries through to this extension,
and thus the two representations are still physically equivalent. 
We will pursue the free-$\A^0$
representation in this paper since it is more elegant, using the existence
of the equivalent *bracket representation to justify its consistency.
(Let us also mention that we checked that just as with normal gauge 
transformations, these supergauge transformations leave the na\"\i ve 
functional measure invariant.)

Fortunately, in contrast to the case for $\A$, 
both the $\sig3$ and $\one$ components of $\C$
are dynamical since they both appear in the kinetic term and $\str\,\C^n$
interactions. In fact they propagate into each other: their only kinetic
term being
\be
\label{c3c0ke}
2N\partial_\mu\C^0\partial_\mu\C^{\sigma}.
\ee
Similarly to the case of $\A_\mu$ we have the option also to consider 
separately the invariant $\C^{\sigma}=\str\,\C/2N$. Again similarly,
under gauge transformations \eq{cgaugetr}, $\C^{\sigma}$ mixes into the other 
components, and thus all components of $\C$ would have the same wavefunction
renormalization if this were needed. Whilst this time the theory does not
itself constrain $\C^{\sigma}$, at first sight it appears that we are 
able to impose the linear gauge invariant constraint:
\be
\label{3constraint}
\str\,\C=2N\Lambda^{D/2-1}.
\ee
(Note again that by \eq{break}, $\C^{\sigma}=\Lambda^{D/2-1}$ 
must be non-zero.)
In contrast to a non-linear constraint we might expect this to leave
the renormalizability or finiteness of the theory undisturbed.

In fact \eq{3constraint} spells trouble since the Lagrangian must also
include a potential $\str\,V(\C)$.
(Such a potential will be used to induce spontaneous symmetry 
breaking and give all remaining `Higgs'' a mass.)
This is because \eq{3constraint} causes \eq{c3c0ke} to vanish.
$\C^0$ thus becomes a Lagrange multiplier imposing
its equations of motion as a constraint:
\be
\label{0constraint}
{\partial\over\partial\C^0}\,\str\,V(\C)=0.
\ee
For a simple mass term $\sim\Lambda^2\str\,\C^2$
(which would be needed to give the $\C^A$ masses of order $\Lambda$)
this constraint leads to the contradiction $\C^{\sigma}=0$.
For the simplest allowable non-trivial $V$, \ie of rank 4, $\C^0$ appears as 
a cubic and \eq{0constraint} sets 
$\C^0$ equal to the roots of a quadratic, with coefficients polynomial
in $\C^A$. As well as being messy this does not look promising for
furnishing a finite theory. For these reasons we do not pursue this
option further. 

Finally, we discuss the form of the Faddeev-Popov ghosts and BRST algebra,
which will appear upon gauge fixing. We write these super-ghosts as 
\bea
\label{ghost}
{\eta} = \left( \begin{array}{cc} \eta^1 & \phi \\
					 \psi & \eta^2  
	       \end{array}
       \right). 
\eea 
When quantizing a bosonic gauge group we introduce fermionic ghosts
so that the ghost action yields the Faddeev-Popov determinant and not
its inverse. Na\"\i vely here we would expect similarly to assign opposite 
grading to $\eta$ so that $\eta^i$ in \eq{ghost} are bosons and $\phi$ and 
$\psi$ are fermions. However, full superfields are of indeterminate grading 
and the usual requirement
of (anti)commutativity is here replaced (for supercoloured objects) 
by the (anti)cyclicity of the supertrace. 
One can readily check that with the above ghost assignments
$\str\,\eta X=-\str\, X\eta$ if $X$ has odd
ghost number, as required, but that 
$\str\,\eta X=\str\,X\sig3\eta\sig3$ if $X$ has even ghost number. 

Even if such cyclicity breaking terms are excluded from the action (by \eg 
being single supertrace and 
total ghost number zero), they can arise in multiple-supertrace terms at one
loop and higher loops. This 
in turn leads to
a breakdown of the (unfixed) global $SU(N|N)$ invariance 
[because the supertrace of a commutator no longer 
vanishes as required \cf \eq{cycle} or \eq{gaugecy}]
and thus presumably spurious $\sig3$ insertions appearing 
in the loop corrections.
Further problems are uncovered when we try to construct the BRST invariance.
Proceeding in standard fashion, we replace $\omega$ by ($\gt$ times) the
ghost in \eq{gaugetr}:
\be
\label{brst0}
\delta \A_\mu=\epsilont[\nabla_\mu,\eta],
\ee
and restore the grading by introducing 
the scalar fermionic parameter $\epsilont$. 
One can readily check that the usual property that
$\epsilont$ commutes with gauge fields and anticommutes with ghosts
is used to prove BRST invariance. However, $\epsilont$ has no simple
(anti)commutation properties with the superfields. Indeed it matters
whether we place $\epsilont$ before or after $\nabla_\mu$, and
$\eta$, and the expressions differ by more than just a sign.

There is an elegant solution:
recall that it is actually a matter of convention
whether different fermionic flavours commute or anticommute \cite{comfermi}.
In other words, we are free to introduce a multiple grading. We will
assign both a supergroup grading f
and a ghost grading g. All superfields
including the ghosts have supergroup-odd block-off-diagonal entries (f $=1$)
and supergroup-even block-diagonal entries (f $=0$). 
$\epsilont$ is supergroup-even: f$(\epsilont)=0$.
$\A$ and $\C$ are ghost-even (g $=0$) and $\eta$, $\etab$ and BRST parameter
$\epsilont$ are ghost-odd (g $=1$). The algebra is completely determined by
the requirement that elements commute up to a multiplicative extra sign
whenever odd elements from the same grading are 
pushed passed each other, \ie for elements $a$ and $b$:
\be
ab=ba (-1)^{\f(a)\f(b)+\g(a)\g(b)}.
\ee
One can readily check that (anti)cyclicity is now preserved \viz
\be
\str\,\eta X=(-1)^{\g(X)}\,\str\,X\eta.
\ee
$\epsilont$ now simply 
(anti)commutes with (ghost-)superfields, and thus the usual form for
the BRST algebra results. This can be used
to prove all the usual properties of gauge fixing \eg independence 
of the choice of gauge, transversality of on-shell Green functions and so 
on, and must thus yield the correct form for the Faddeev-Popov 
superdeterminant. In the next section we will give the explicit form of
the gauge fixing function and ghost action that we will use, 
and in \sec{Ward} the explicit form of the corresponding BRST algebra.

\section{Spontaneously broken action}
\label{Spontaneously}

Having settled the issues specific to the choice of $SU(N|N)$, we
now describe the full construction. 
Let the super-gauge field of \SUNN \  be denoted by $\Amu \equiv
\A^\alpha_\mu S_\alpha$. We can write
this in supermatrix form with  bosonic diagonal elements and
fermionic off-diagonal elements:
\be
\label{defA}
\Amu = \left( \begin{array}{cc} A^1_\mu & B_{\mu} \\
			       \Bbar_{\mu} & A^2_\mu    
	       \end{array}
       \right) + \A^0_\mu\one.
\ee
As discussed in sec. \ref{Alternatives}, $\A^{0}_\mu$ does not actually 
appear in the action, and we may either leave it in the theory where it has 
no effect (as we do here) or as shown in \sec{Alternatives}, define it away 
entirely by modifying the Lie bracket selectively. Once again,
our covariant derivative is taken\footnote{Unlike
refs. \cite{manerg,erg2,erg1} rescaling $\Amu \rightarrow \Amu/\gt$ will be
of no benefit here as we will fix the gauge.} to be $\nabla_{\mu} =
\partial_{\mu} - i\gt\Amu$ with the field strength being ${\cal F}_{\mu\nu}:=
{i \over \gt}[\nabla_{\mu},\nabla_{\nu}]$.  However it will prove helpful
to make explicit the scale $\Lambda$ hidden in the coupling
constants when the number of (Euclidean) 
dimensions $D\ne4$, by writing $\gt=g\Lambda^{2-D/2}$, from now on.
For the sake of generality,
we introduce the covariant higher derivatives via functions $W$.
We introduce the convenient notation
\be
\label{ourchoice}
{ u}\{W\}{ v}={ v}\{W\}{ u} = 
\str\! \dint\, {u}(x)W(-\nabla^2/\Lam^2)\cdot{v}(x),
\ee 
taking $W\cdot {v}$ 
to mean that each $\nabla_{\mu}$ acts via
commutation. Let $\c$ be a polynomial of rank $r$.
We can then write the pure Yang-Mills part of the action as 
\be\label{YMact}
S_{YM} = {1\over {2}}{\cal F}_{\mu \nu} \{\c\}{\cal
F}_{\mu \nu}
\ee
[\cf eq. (\ref{p-cp})].

Actually, these objects naturally arise and have a deeper meaning
within the exact renormalization
group \cite{catchall,manerg,erg1,erg2}. Thus $c$ is a cutoff function,
and there is actually a wide choice of the exact form of covariantization 
$\{W\}$. The formalism we present is independent of the choice \eq{ourchoice}
except for some particulars of the power counting proof in \sec{Counting}.

The super-scalar field $\C$ is introduced:
\be
\label{defC}
\C = \left( \begin{array}{cc} C^1 & D \\
			      \Dbar & C^2       
	       \end{array}
       \right),
\ee
with no restriction  placed upon $\C$. Thus $\C$ can be expressed uniquely
in terms of components as \eq{Cexp}, or more simply just as an unconstrained
Hermitian supermatrix field with components $\C^i_{\gap j}$. We require 
that classically $\C$ picks up an expectation value 
$<\!\C\!>\, = \sig3 \Lambda^{D/2-1}$, 
so that \SUNN \  is spontaneously broken to the bosonic subgroup 
$SU(N) \times SU(N)$. (Again, the
na\"\i vely expected $U(1)$ associated to $\A^0$ has no effect or 
does not appear, \cf \sec{Alternatives}.)
The contribution of the $\C$ field to the action is chosen to be  
\be\label{unbCact}
S_{\C} =
{1\over2}\nabla_{\mu}\!\cdot \C \{\ctil \}\nabla_{\mu}\!\cdot \C  + 
\frac{\lambda}{4}\Lambda^{4-D}\,
\str\!\int\!d^{D}x\,\left({\cal C}^{2} - \Lambda^{D-2}
\right)^2.
\ee
Note the introduction of a new cutoff function ${\tilde c}$. Again it
is convenient to choose $\ctil$ to be a polynomial, this time of rank $\rtil$. 
By construction $\C=\sig3\Lambda^{D/2-1}$ is a stationary point
of the potential. Expanding about this (\ie
$\C \mapsto \C+\sig3\Lambda^{D/2-1}$), the action (\ref{unbCact}) becomes
\bea\label{brkCact}
\lefteqn {S_{\C} = 
-{g^2\over2}\Lam^2[\Amu,\sig3]\{\ctil\}[\A_{\mu},\sig3] 
-ig\Lam [\Amu,\sig3]\{\ctil\}\nabla_{\mu}\!\cdot \C }
\nonumber \\
& & \hspace{1cm} 
+{1\over2}\nabla_{\mu}\!\cdot \C \{\ctil \}\nabla_{\mu}\!\cdot \C
+ \frac{\lambda}{4}\Lambda^{4-D} \,\str\!\int\!d^{D}x\, \left(
\Lambda^{D/2-1} \{ \sigma_{3},{\cal C}\} + {\cal C}^{2}\right)^2.
\eea
The first term of (\ref{brkCact}) gives a mass of order the effective
cutoff to the fermionic part of $\A$. The bosonic part of $\C$ also gains a
mass via the last part of (\ref{brkCact}). 

To further investigate the properties of this action, we need to fix 
the gauge. To rid us of the part in the second term of (\ref{brkCact}) that
mixes  single powers of $\A$ and $\C$ fields, we follow 't Hooft's lead
\cite{thooft} and make the following choice of gauge fixing function:
\be
\label{gafif}
{F} = \partial_{\mu}\A_{\mu} - ig{\Lam \over 2\xi} {{\hat c} \over
{\tilde c}}[\sig3,\C], 
\ee
utilising another new cutoff function ${\hat c}$, $\chat$ being chosen
polynomial of rank $\rhat$. Here the cutoff functions have argument
$(-\partial^2/\Lam^2)$ because, being part of the gauge fixing, there is no
need for covariantization in \eq{gafif}. After 't Hooft averaging, the 
gauge fixing part of the action is
\bea\label{gaugeact}
\lefteqn
{S_{Gauge} =  
\xi\,  {F}\cdot \chat\! \cdot {F}  \nonumber } \\
& & \hspace{0.6cm} =  
\xi\, (\partial_{\mu}\A_{\mu})\cdot\chat\!\cdot(\partial_{\nu}\A_{\nu})
- ig\Lam\,(\partial_{\mu}\A_{\mu})\cdot\ctil\!\cdot[\sig3,\C] 
\nonumber  \\
& & \hspace{1.5cm}   \,
- g^2{\Lam^2 \over 4\xi}[\sig3,\C]\cdot{{\hat c}\over{\tilde c}^2}
\cdot[\sig3,\C] ,
\eea
using the notation 
${\ds { u}\cdot W\cdot {v} \equiv \str \dint \,{
u}(x)\,W(-\partial^2/\Lam^2)\,{v}(y)}$. Introducing (\ref{gaugeact}) 
into the action  cancels the required term as well as providing a mass 
term for the fermionic part of $\C$. 

Apart from the two $SU(N)$ gauge fields (and the decoupled or missing 
$\A^0_\mu$), all fields have masses of order 
the cutoff $\Lambda$. Note that in the usual unitary gauge interpretation, 
the fermionic part of $\C$ is the would-be Goldstone mode which is
eaten by the fermionic part of $\A$. $A^1_\mu$ is the $SU(N)$ field we
set out to regulate. $A^2_\mu$ is unphysical because the sign of its
action is the opposite of $A^1$ by the supertrace (\ref{supertrace1}).
As we explain in sec. \ref{Unitarity}, this leads to unitarity violations.
Fortunately, since the two gauge fields belong to two different $SU(N)$
groups there is no bare interaction between them. Indeed 
any such interaction would have to involve products of gauge 
invariant supertraces. In sec. \ref{Unitarity}, we use this
insight to show that the two sectors must decouple in the limit that 
the cutoff is removed.

The gauge fixing also introduces the Faddeev-Popov ghost super-fields 
\eq{ghost}. To tidy up the contribution to the action, we change antighost
variables: $\bar{\eta} \rightarrow \chat \tilde{c} \, \bar{\eta}$. 
As we will see in sec. \ref{Counting}, this has the added benefit of 
ensuring that power counting arguments will be assigning the correct 
momentum behaviour to some of the ghost
interaction vertices. Consequently, the ghosts  appear in the action as
\bea\label{ghostact}
S_{Ghost} = 
-2 {\bar\eta} \cdot \chat \tilde{c}\cdot \partial_{\mu}\nabla_{\mu}
\!\!\cdot\!\eta - \frac{g^2}{\xi}\,
\str\!\int\! d^{D}\!x\;  [\sigma_{3}, \bar{\eta}] 
 [\Lam^2\sigma_{3}+\C\Lambda^{3-D/2},\eta].
\eea
As we saw in \sec{Alternatives}, the introduction of a separate ghost
grading ensures that simply an extra sign appears whenever two ghosts
are moved passed each other. 

Finally, in order to keep the high momentum behaviour of the $\A$ propagator
unchanged by the introduction of the $\C$ field and gauge fixing, we
require the  ranks of our polynomial cutoff functions to be bounded as 
$\rhat \geq r > \rtil -1$. In the next section, in
order to get proper bounds on these indices, it will be convenient to take
$r$, $\rtil$, $\rhat$  as general real
numbers, the restriction to integers being taken at the end. As a matter of
fact it is consistent to take these parameters to be real having in mind 
more general cutoff functions (analytic around the origin, $p=0$, and with
asymptotic behaviour $c^{-1} \sim
{p^{2r} \over \Lambda^{2r}}$ \etc). In this case strictly speaking we should
add the condition $\rtil>-1$ which is necessary to ensure that 
the high momentum behaviour of the $\C$ field is unaffected by the 
spontaneous symmetry breaking mass term in \eq{brkCact}. Thus the following
conditions are required on the indices:
\be
\label{behav}
\rhat \geq r > \rtil -1\qquad{\rm and}\qquad \rtil>-1.
\ee

\section {Power counting}
\label{Counting}

We now establish the finiteness of this theory, to all orders in $\hbar$.
We start by computing an upper bound on
the superficial degree of divergence of any one-particle-irreducible (1PI) 
diagram and show that this is negative in any dimension $D$,
for all but a small number
of one-loop diagrams, providing the indices $r$, $\rtil$ and $\rhat$
satisfy the inequalities \eq{finalcons}. 
Then in secs. \ref{Supertrace}
and \ref{Ward}, we establish that these one-loop exceptions are themselves
finite in all dimensions $D<8$,
as a consequence of cancellations resulting from the supersymmetry 
of $SU(N|N)$ and gauge invariance. Since the superficial degree of
divergence of any given diagram and all its connected proper sub-diagrams
is thus shown to be negative, finiteness to all loops follows
from the convergence theorem \cite{itz}.

Using standard rules for calculating the superficial degree of
divergence \cite{itz} of a 1PI diagram in $D$
space-time dimensions, we get
\be \label{sdd1}
\begin{array}{l}
{\ds {\cal D}_{\Gamma} = D L - (2r+2) \, I_{\A} -(2\rtil+2) \, I_{\C}
-(2 \rhat -2\rtil+2) \, I_{\eta}  + \sum_{i=3}^{2r+4} (2r+4-i) \,
V_{{\A}^i} }\nonumber\\
{\ds + \sum_{j=2}^{2\rtil+2} (2\rtil+2-j) \, V_{{\A}^j {\C}}
+ \sum_{k=1}^{2\rtil+2} (2\rtil+2-k) \,
V_{{\A}^k {\C}^2} + (2 \rhat -2\rtil+1) \, V_{{\eta}^2 \A}, }
\end{array}
\ee
where $L$ is the number of loops and $I_{f}$ and $V_{f}$ correspond to the
number of internal lines and vertices of $f$-type respectively. In 
\eq{sdd1}, inequalities \eq{behav} 
have already been
assumed for the degree of divergence of the vector and $\C$ propagators
respectively to be counted properly. 

As it stands, \eq{sdd1} does not account properly for 1PI diagrams with
external anti-ghost lines. In fact, the whole momentum dependence of the
$V_{{\eta}^2 \A}$ vertex 
is counted as flowing in the loop(s), without taking into account the
fact that such a
dependence is actually only carried by $\bar{\eta}$ lines and, thus, that one has to
check whether such lines are external or not. This results in a systematic
overestimate of \dga. In order to
remedy this and, thus, improve our upper bound, \dga, we add
$-(2 \rhat-2\rtil+1) E_{\bar{\eta}}^{\A}$, with $E_{\bar{\eta}}^{\A}$
being the number of external anti-ghost lines which enter $V_{{\eta}^2 \A}$
vertices. Therefore, the improved formula for the superficial degree of 
divergence is   
\be \label{sdd1im}
\begin{array}{l}
{\ds {\cal D}_{\Gamma} = D L - (2r+2) \, I_{\A} -(2\rtil+2) \, I_{\C}
-(2\rhat-2\rtil+2) \, I_{\eta}  + \sum_i (2r+4-i) \, V_{{\A}^i}  }\nonumber\\
{\ds +\sum_j (2\rtil+2-j) \, V_{{\A}^j {\C}}
+ \sum_{k} (2\rtil+2-k) \,
V_{{\A}^k {\C}^2} + (2\rhat-2\rtil+1) \left( V_{{\eta}^2 \A} - E_{\bar{\eta}}^{\A} \right). }
\end{array}
\ee
The variables upon which ${\cal D}_{\Gamma}$ is dependent can be easily
related to the number of external lines of each type, $E_{f}$, as
\bea 
L &=& 1 + I_{\A} + I_{\C} + I_{\eta}\nonumber\\
 &{}&\quad - \sum_i
V_{{\A}^i} - \sum_j V_{{\A}^j {\C}}
- \sum_{k} V_{{\A}^k {\C}^2} -
V_{{\eta}^2 \A} - V_{{\eta}^2 \C} - V_{{\C}^3} - V_{{\C}^4}, 
\label{lrel}\\
E_{\A} &=& -2 I_{\A} +  \sum_i
i V_{{\A}^i} + \sum_j j V_{{\A}^j {\C}} +
\sum_{k} k V_{{\A}^k {\C}^2} +
V_{{\eta}^2 \A}, \label{earel}\\
E_{\C} &=& -2 I_{\C} + \sum_j V_{{\A}^j {\C}} +
2 \, \sum_{k} V_{{\A}^k {\C}^2} +
3 V_{{\C}^3} + 4 V_{{\C}^4} + V_{{\eta}^2 \C}, \label{ecrel}\\
E_{\eta} &=& E_{{\eta}}^{\A} + E_{{\eta}}^{\C} +
E_{\bar{\eta}}^{\A} + E_{\bar{\eta}}^{\C} = -2 I_{\eta} + 2 V_{{\eta}^2 \A}
+ 2 V_{{\eta}^2 \C}. \label{eerel}
\eea
In the last of the above relations, to ensure consistency with previous
notation we split external ghost lines according to the vertices they are 
attached to. Thus $E_{\eta}^{f}$ ($E_{\bar\eta}^{f}$), 
$f \!=\! \A,\C$, is the number of external 
(anti)ghost lines entering $V_{{\eta}^2 f}$ vertices; they satisfy
the expected constraint $E_{{\eta}}^{\A} + E_{{\eta}}^{\C} =
E_{\bar{\eta}}^{\A} + E_{\bar{\eta}}^{\C}$.
We note for later that \eq{eerel} may thus be written
\be
\label{alteerel}
E^\A_\etab+E^\C_\etab=-I_\eta+V_{\eta^2\A}+V_{\eta^2\C},
\ee
as can be most simply understood by deriving the 
equation directly as a count over external antighosts.
\eq{lrel} is valid for connected diagrams only, as
the first term in the r.h.s. - representing the number of connected
components - has been set to $1$.

By making use of the above formulae, it is possible to rewrite
${\cal D}_{\Gamma}$ in a more useful form, independent of internal lines,
\be \label{sdd2}
\begin{array}{l}
{\ds {\cal D}_{\Gamma} = (D-2r-4) \, (L-2) - E_{\A} -(r-\rtil+1) \,
E_{\C} -2 (r+\rtil-\rhat +1) \,
E_{\bar{\eta}}^{\C} \phantom{\sum_{j=1}^2} } \nonumber\\
{\ds - (2r+3) \, E_{\bar{\eta}}^{\A} - (r -\rtil + 1) \sum_j
V_{{\A}^j {\C}} + (r-3\rtil-1) \, V_{{\C}^3}  + 2 (r-2\rtil) \,
V_{{\C}^4} }\nonumber\\
{\ds + (r+\rtil-2\rhat -1) \, V_{{\eta}^2 \C} + 2 (D-r-2). }
\end{array}
\ee


We now establish necessary and sufficient constraints on 
$r$, $\rtil$ and $\rhat$, such that ${\cal D}_{\Gamma}$ is
negative for as many diagrams as possible.
Not all diagrams can be regularised this way. For example, the superficial 
degree of divergence of the one-loop diagrams involving only $\A$ fields
is $D - E_{\A}$, which is non-negative for $E_\A\le D$
independent of the parameters $r$, $\rtil$ and $\rhat$.
We will start with the proof of a proposition which will help us dispense
with some of the constraints we find; then we will analyse diagrams 
with two or more loops 
and, after, we will return to one-loop
diagrams. 

Let us denote by $\S$
the collection of triples $(r,\rtil,\rhat)$ such that \dga \, is negative for
any given set of 1PI diagrams and \eq{behav} holds.\\[0.3cm]
{\bf Proposition 1}: {\sl
If $(r_0,\rtilo,\rhato) \in \S$, then the subset
$\big\{ (r, \rtil, \rhat) \;\, s.t. \,\; r \!\geq \! r_0, \, \rtil \! = \!
\rtilo, \, \rhat \! \geq \!
\rhato,\\ \rtilo -1 \! < \! r \! \leq \! \rhat \big\} \subset
\S$. }\\[0.3cm]
{\sl Proof:}\\[0.3cm]
The proof is essentially based on the one-particle-irreducibility of
diagrams.

The whole dependence of \eq{sdd2} on $\rhat$ amounts to $ 2 \rhat \, \big(
E_{\bar{\eta}}^{\C} -
V_{{\eta}^2 \C} \big)$, which is always non-positive as it is not possible to
have more external anti-ghost lines entering $V_{{\eta}^2 \C}$ vertices
than $V_{{\eta}^2 \C}$ vertices themselves. Thus, increasing $\rhat$ above
$\rhato$ can only leave stationary or decrease an already negative \dga.

As far as $r$ is concerned, it enters \eq{sdd2} as
\be
r \Big( -2L + 2 - E_{\C} -2
E_{\bar{\eta}}^{\C} - 2 E_{\bar{\eta}}^{\A} - \sum_j
V_{{\A}^j {\C}} + V_{{\C}^3}  + 2 \, V_{{\C}^4} + V_{{\eta}^2 \C} \Big) = 2
r \Big( \sum_i V_{{\A}^i} - I_{\A} \Big),
\ee
where the last equality follows by using \eq{lrel}--\eq{eerel}, or directly
from \eq{sdd1im}. This contribution is always non-positive as we know that 
in a 1PI diagram every
$V_{{\A}^i}$ vertex must attach to at least two internal $\A$ lines.
Therefore increasing $r$ above $r_0$ can at most cause \dga \, to decrease
further. \hfill $\Box$

\subsection{Multiloop graph analysis}

In order for every possible $L\ge2$ loop 1PI diagram to have a negative 
\dga, we can impose all coefficients in \eq{sdd2} to be negative and, thus, 
get sufficient conditions. This amounts to the following relations
\be \label{conditions}
r>D-2, \quad \qquad 2r+3>0, \quad \qquad r<2\rtil , \quad \qquad
 \rhat<r+\rtil+1, 
 \ee
together with \eq{behav}. ({\it N.B.} In the case of polynomial cutoff 
functions, it is easy to see that there are integers $r$,
$\rtil$, $\rhat$ satisfying \eq{behav} and \eq{conditions}.)

The conditions (\ref{conditions}) imply a lower bound on $\rtil$,
$\rtil> {1\over 2} \max \left(D-2, -{3\over 2}\right)$, as well. 
The $D$-dependent lower bounds
on $r$ and $\rtil$ may be expected to be also necessary, as the higher
the space-time dimension, the more divergent the diagrams.
However, physics does not provide any reasonable arguments to
explain {\emph{upper}} bounds on $\rhat$ and $r$, apart from $r \leq \rhat$
[\cf \eq{behav}]. In fact they are not necessary, as one can easily
appreciate by making use of Proposition 1: applying it to all 
triples $(r_0,\rtilo,\rhato)$ which
satisfy \eq{conditions} and \eq{behav}, we see that the 
third and the fourth
inequalities in \eq{conditions} are not necessary,
and we are thus left with the sufficient relations 
\be
\label{cons}
r>\max \left( D-2, -{3\over 2} \right),\qquad \rtil>\max \left( {D \over
2}-1, -{3\over 4} \right) \qquad {\rm and}\qquad 
\rhat \geq r > \rtil-1.
\ee 
If we only rely on power counting, these conditions are also necessary, 
for any $D \geq {1\over 2}$, as they ensure finiteness in the two
two-loop vacuum diagrams with only $\A^3$ and $\C^4$ vertices
respectively. Actually, both these diagrams vanish by the supertrace
mechanism explained in \sec{Supertrace}, but we will see that there
are non-vanishing one-loop diagrams that require the same conditions.

Before moving to one-loop diagrams, it may be helpful to illustrate the use
of Proposition 1 within a restricted class of multiloop graphs where it
easier to see in detail what is going on.
Let us focus on the subset of multiloop {\sl vacuum} diagrams formed from only 
$\C^4$ vertices. The superficial degree of
divergence takes a very simple form, ${\cal D}_\Gamma = (D-2r-4) \, (L-2) + 2
(r-2\rtil) \, V_{{\C}^4} + 2 (D-r-2)$, which is negative for every possible
diagram in the $L\ge2$ set
provided the relations $r>D-2, r <2\rtil$ are imposed. Again, these
conditions imply a lower bound on $\rtil$, $\rtil>{D \over 2}-1$. 
As the number of $\C^4$ vertices can be arbitrarily large, one can be
misled and conclude that the relation $r<2\rtil$ is also necessary as an
asymptotic condition. However, increasing the number of $\C^4$ interactions 
also increases the number of loops, as follows for this restricted
set from eqs. \eq{lrel}--\eq{eerel}. Indeed in this simple case, 
reexpressing $L$ in terms of $V_{{\C}^4}$ 
yields ${\cal D}_\Gamma = (D-4\rtil-4) \, (V_{{\C}^4} -1) + 2
(D-2\rtil-2)$. \dga \ is thus independent of $r$, hence increasing $r$ does not
change it, which is actually the essence of Proposition 1.
One is then left with the (necessary) constraint $D-2\rtil-2<0$. 
Whilst for a number of general classes of diagrams, 
we can similarly demonstrate that \eq{cons} provides
sufficient conditions by reexpressing \dga\ using \eq{lrel}--\eq{eerel}, 
this is
not possible for the full set of multiloop graphs. Fortunately Proposition
1 comes elegantly to the rescue.


\subsection{One-loop diagram analysis}

As mentioned below \eq{sdd2}, not all one-loop diagrams can be
regularised by imposing constraints on the ranks of the cutoff functions. This
is why several sub-cases are to be analysed
when dealing with one-loop graphs.
Nonetheless, the strategy we are going to use is pretty much the
same as in the previous subsection. 
We will first rewrite \dga \ in terms of the proper, non-negative variables
and, then, we will impose
all its coefficients to be negative so as to get sufficient
conditions. Finally, we will relax some of those conditions by means of 
Proposition 1, and prove the remainder to be necessary.

Let us start with specialising \eq{sdd2} to $L=1$: 
\be
\label{sdd1loop}
\begin{array}{l}
{\ds \dgaone = D - E_{\A} -(r-\rtil+1) \,
E_{\C} -2 (r+\rtil-\rhat +1) \, E_{\bar{\eta}}^{\C} - (2r+3)\, 
E_{\bar{\eta}}^{\A} \phantom{\sum_{j=1}^2} } \nonumber\\
{\ds - (r -\rtil + 1) \sum_j
V_{{\A}^j {\C}} + (r-3\rtil-1) \, V_{{\C}^3}  + 2 (r-2\rtil) \,
V_{{\C}^4} + (r+\rtil-2\rhat -1) \, V_{{\eta}^2 \C}. }
\end{array}
\ee

The first set of diagrams we are going to analyse consists of all 
the one-loop diagrams with {at least} $D+1$ external $\A$ lines - so
that the combination $(E_\A -D-1)$ is always non-negative within the set -  
and {any} number of external $\C$  and (anti-)ghost lines.
Reexpressing \eq{sdd1loop} in terms of $(E_\A -D-1)$ amounts
to replacing $D-E_\A$ with $-(E_\A -D-1)-1$ without altering the rest of
the expression.
Therefore, in order for all its coefficients 
to be negative, we have to impose  
\be
\label{1loopconditions}
r<2\rtil, \quad \qquad 2r+3>0, \quad \qquad \rhat<r+\rtil+1,
\ee
together with \eq{behav}.
 
The next set we will deal with is made up of all the one-loop diagrams
with {at least} one external antighost line, $E_{\bar{\eta}}^{\A,
\C} \geq 1$, and {any} number of external $\A$'s  and $\C$'s.

In the case of an $\A$-type antighost line, upon changing
$E_{\bar{\eta}}^{\A}$ to $(E_{\bar{\eta}}^{\A}-1)$ we get
$\dgaone = (D-2r-3) - (2r+3) \, (E_{\bar{\eta}}^{\A} -1) + \cdots$, where
the ellipsis stands for unchanged terms in \eq{sdd1loop}.
Demanding that all coefficients in this relation be negative results in 
the extra condition $D-2r-3<0$ [together with \eq{1loopconditions}]. 

In the case of a $\C$-type antighost line, we get a different 
constraint. Introducing $(E_{\bar{\eta}}^{\C}-1)$ 
into \eq{sdd1loop} causes it to change to $\dgaone = D -2 (r+\rtil-\rhat
+1) -2 (r+\rtil-\rhat +1) \, (E_{\bar{\eta}}^{\C} -1) +
\cdots$, where again unchanged terms have not been written down. 
 Imposing the coefficients in the above expression to be negative 
yields the condition $ D -2(r+\rtil-\rhat +1) <0$, which has to replace 
the upper bound on $\rhat$ in \eq{1loopconditions} for any $D \geq 0$.

Let us now analyse the set of one-loop graphs with {at least} two external
$\C$'s and {any} number of external $\A$  and (anti-)ghost lines. 
Rewriting \eq{sdd1loop}
as $\dgaone = D -2 (r-\rtil+1) - (r-\rtil+1) \,
(E_{\C}-2)+ \cdots$ and demanding all its coefficients to be negative we
find another constraint, $r-\rtil>{D \over 2}-1$, which will turn out to
be necessary as well.

Finally, let us take into account one-loop diagrams with {just one}
external $\C$ and {no} external $\A$  nor (anti)ghost lines. We need not
consider the more general case when any number of external
$\A$'s and (anti)ghosts is allowed because, within the ranges 
\eq{1loopconditions} and \eq{behav}, they both contribute negatively to
$\dgaone$, irrespective of the choice of vertices. 
Despite containing three elements only, depending on the `flavour'
$\A$, $\eta$ or $\C$, of the
loop, this set of diagrams gives us two further conditions: $r-\rtil>{D 
\over2}-1$ and $\rhat-\rtil>{D \over 2}-1$ 
for the $\A$ and $\eta$ loop flavours respectively.
The latter, however, is always fulfilled if the former is, $\rhat$ being
greater than or equal to $r$. 

The results of one-loop diagram analysis can be therefore summarised as
follows: all the $L\!=\!1$ diagrams except those 
with up to $D$ external $\A$ legs and no $\C$  or $\etab$ (and thus also 
$\eta$) external lines, are regulated by imposing the following constraints,
\be
\label{1looprels}
\begin{array}{l}
{\ds 2r+3>0, \quad \qquad \rhat<r+\rtil+1, \quad \qquad D-2r-3<0, }\nonumber\\[0.25cm]
{\ds r<2\rtil, \quad \qquad \rhat<r+\rtil +1 -{D\over 2}, \quad \qquad r-\rtil >
{D\over 2}-1, 
} \nonumber\\  
\end{array}
\ee
together with \eq{behav}.

By inspection of \eq{sdd1loop} it is possible to reduce further the set of
diagrams that appear to remain unregularised: within the set of one-loop 
diagrams with up to $D$ external $\A$ legs and no $\C$  or (anti)ghost 
external lines, only those formed from
$V_{\A^k \C^2}$, $V_{\A^i}$, $V_{\eta^2 \A}$ need to be analysed. 
(These vertices contribute nothing to $\dgaone$.
From topological considerations $V_{\C^3}=V_{\C^4}=V_{\eta^2\C}=0$
and either $\sum_j V_{\A^j\C}=0$ or $\sum_j V_{\A^j\C}\ge2$ in which
case by \eq{1looprels} already $\dgaone<0$.)
It is helpful to give this set a name:
the  ``One-loop Remainder Diagrams''. 
     
Relations \eq{1looprels} imply lower bounds on $\rtil$,
$\rtil> {1\over 2} \max
\left( D-2, {{D-3}\over 2}, -{3\over 2} \right)$, and on 
$r$, $r>\max \left( D-2, {{D-3}\over 2}, -{3\over 2} \right)$, as well. To get 
the former note that $r<\rhat<r+\rtil +1 -{D\over 2}$ and use $2r+3>0$,
$D-2r-3<0$; as for the latter, use $r-\rtil > {D\over 2}-1$ together with
$2r+3>0$, $D-2r-3<0$. 

By making use of Proposition 1 it is possible to rid us of the upper
bounds on $r, \rhat$, so that we are left with
\be
\label{1looprels2}
r> \max \left( D-2, {{D-3}\over 2}, -{3\over 2} \right), \quad \rtil>
{1\over 2} \max \left( D-2, {{D-3}\over 2}, -{3\over 2} \right),  \quad
r-\rtil > {D\over 2}-1 
\ee
and \eq{behav}.

For any $D\geq 1$ the above set of solutions contains \eq{cons},
together with one new relation, $r-\rtil>{D\over 2}-1$. 
We have already seen that these conditions are also necessary for 
regularisation purely by power counting, \ie if we 
ignore cancellations arising from the supertrace mechanism. In fact
there are one-loop diagrams for which 
these conditions are necessary even with 
the supergroup factors taken into account. To see this, we borrow a 
result from the next section, that group theory factors for
unbroken one-loop corrections take the form of
a product of two supertraces over the external fields (resulting
in vanishing terms whenever one supertrace is empty or contains
only a single $\A$). It follows that in the broken $SU(N|N)$ theory, the 
results still take this form except that $<\!\C\!>\, = \sig3 
\Lambda^{D/2-1}$ terms may also appear in the supertraces.
Now, the condition $r-\rtil>{D\over 2}-1$
arose from power counting the one-loop graph made by attaching an $\A$
propagator to the $\C\A^2$ vertex [\ie by inspection the vertex from 
$-ig\Lam [\Amu,\sig3]\{\ctil\}\nabla_{\mu}\!\cdot \C$ of \eq{brkCact}].
Thus $r-\rtil>{D\over 2}-1$ is necessary for the contributions 
with group theory factor $\str\,\C\,\str\,\sig3$ (which one can readily
check are non-vanishing).
The condition $\rt>D/2-1$ is necessary for 
finiteness of $(\str\,\C)^2$ contributions arising from attaching a $\C$ 
propagator to the $\str\,\C^4$ vertex, or for the $\str\,\C\,\str\,\sig3$
term arising from attaching the $\C$ propagator to the $\C^2\{\C,\sig3\}$
vertex. (Again one can confirm that these contributions are non-vanishing.)
The final condition for any
$D\ge1$, namely $r>D-2$, already follows from combining these two.

The analysis of one-loop diagrams can also be performed by adopting a
completely different strategy, based on a form of {\it
divide and conquer} algorithm. Here we will explain only the general idea,
without going into details \cite{rome}.   
We start with cutting up diagrams into tadpole-like pieces, 
defined as the sub-diagrams which contain just one internal propagator 
attached to one vertex \cite{erg2}. This can be done in two different ways,
according to which propagator remains attached to the vertex being cut. 
We then compute the degree of divergence of every possible piece
we can end up with, aiming to show that
they all contribute negatively to the overall \dga. If this is the case - 
and it is indeed - what is left is just the analysis of the simplest 
possible one-loop graphs, as any other can be obtained by inserting
tadpole-like pieces, which causes \dga \ to decrease further.
In other words, one can
always bound from above the degree of divergence of a one-loop diagram by
removing tadpole-like pieces one by one and joining together the rest of
the diagram - hence increasing the overall
\dga. Eventually one will be left with a very simple graph, usually a proper
tadpole.
The first part of the analysis, that is calculating \dga \ for such
``components'', is straightforward: by inspection of
\eq{sdd1} it is easy to appreciate that all the possible sub-diagrams
contribute negative terms within the bounds we have already set on
$r$, $\rtil$ and $\rhat$ (eqs \eq{behav} and \eq{conditions}). 
However, this is not all that we need: when
sewing back the diagram after a tadpole-like piece has been removed, 
it is possible to end up with a different vertex. This happens
every time the sewn propagator is different from the one removed.  
Such effect should be taken into account as different vertices contribute
differently to \dga.
The second part of the analysis, \ie showing that all the simplest possible
one-loop diagrams\footnote{except those with no external $\C$ or $\etab$
lines and with up to $D$ external $\A$ legs.} can be
regulated by a suitable choice of $r$, $\rtil$ and $\rhat$, is quite long but
straightforward as well.
One further constraint, $r-\rtil > {D\over 2}-1$, is obtained, 
precisely as before.

To summarise, all but the small set of One-loop Remainder 
Diagrams, as defined below \eq{1looprels}, can be regulated simply
by a suitable choice of the ranks of the cutoff functions; from 
\eq{1looprels} and \eq{behav} the allowed ranges for any $D\geq 1$ are 
\be
\label{finalcons}
\rhat\geq r,\qquad r-\rtil>{D\over2}-1,\qquad \rtil>{D\over 2}-1.
\ee

It is interesting and comforting to note that these resulting relations for 
$r$ and $\rt$ are precisely the ones deduced for our earlier 
un-gauge-fixed but more limited regularisation scheme \cite{erg2}.
For the case that the inverse cutoff functions are polynomials,
and $D$ is integer greater than or equal to 2, these inequalities imply
$\rhat \geq r \geq D-1$, 
$\rtil \geq \left[ \frac{D-2}{2} \right] +1$ and $r-\rtil \geq \left[
\frac{D-2}{2} \right] +1$, $[x]$ being the integer part of $x$.

\section {Supertrace mechanism} 
\label{Supertrace}

We have seen that the conditions \eq{finalcons} are
necessary and sufficient to ensure that, in $D\ge1$ dimensions,
all diagrams are superficially
finite already by power counting, with the exception of the One-loop 
Remainder
Diagrams: those formed from only $\C^2\!\A^j$, $\A^j$ and $\etab\A\eta$ 
vertices, with $D$ or less external $\A$ legs and no external $\C$ or 
(anti)ghost legs.
By the power counting \eq{sdd1loop}, their superficial degree of divergence
is $\DG=D-E_\A$.
Actually, these diagrams are also finite as a consequence of cancellations
that are not incorporated in the power counting analysis. In this 
section we show that of these one-loop diagrams, the ones with three 
or less external $\A$ legs are finite as a consequence of the supersymmetry
of the unbroken $SU(N|N)$. This cancellation will be referred to as the 
supertrace mechanism. The One-loop Remainder Diagrams with $3<E_\A\le D$, 
will be shown in sec. 
\ref{Ward} to be finite in all dimensions $D<8$
as a consequence of this 
mechanism plus constraints arising from gauge invariance.
Actually, as a bonus, 
we will see in this section that in all the One-loop Remainder 
Diagrams, the parts arising from the spontaneous
symmetry breaking are already finite by power counting. This observation
will prove useful in simplifying the analysis in sec. \ref{Ward}.

Clearly One-loop Remainder Diagrams are formed in one of three ways: either 
we use only 
$\C$ propagators joining $\C^2\!\A^j$ vertices together, or we use $\A$ 
propagators to join just $\A^j$ vertices together, or finally we can use 
just $\eta$ propagators to join just $\etab\A\eta$ vertices together. It
will prove useful to construct the one-loop diagrams in two steps: by
first constructing a tree diagram and then closing the tree into a loop
by attaching a further propagator. We discuss the $\C$ case first. 

\subsection{One-loop Remainder Diagrams with $\C$ propagators}

For large momentum the $\C$ propagator behaves as: 
\be
\label{cprop}
<\!\C^i_{\gap j}(p)\,\C^k_{\gap l}(-p)\!> \,={\ct\over p^2}
\delta^i_{\gap l}\left(\sig3\right)^k_{\gap j}+O(p^{-4-2m}),
\ee
where $\ct\equiv\ct(p^2/\Lambda^2)$ and $m={\rm min}(2\rt,\rh)$.
The first term on the r.h.s comes from inverting the 
unbroken part of the kinetic term in \eq{brkCact}, and at large momentum
gives the behaviour already incorporated in the power-counting analysis,
and the second term gives the asymptotic behaviour coming from the symmetry 
breaking terms (namely the mass term ${\lambda\over4}\Lambda^2\{\sig3,\C\}^2$
in \eq{brkCact} and the last term in \eq{gaugeact}. The propagators may be 
computed in the usual way by adding a source term to the Lagrangian, 
and noting that the unbroken kinetic term has the 
form ${1\over2}\C^i_{\gap j}[p^2\ctil (\sig3)^l_{\gap i}\delta^j_{\gap k}]
\C^k_{\gap l}$. In sec. \ref{Ward}, it will prove convenient to introduce
the source as $\str\,\J\C$ 
where ${\cal J}$ is a supermatrix field and thus $(\sig3\J)^T$ is the
usual source.) 

We see by \eq{finalcons} that parts of the one-loop integral involving the 
symmetry breaking mass term in \eq{cprop}, are already finite since
their degree of divergence is bounded by $\DG\le D-E_\A-2
-2\,{\rm min}(\rt,\rh-\rt)<0$. 
Thus, since the $\C^2\!\A^j$ vertices come from the
unbroken part of \eq{brkCact}, we note that the potentially divergent
contribution has the same structure as the symmetric $SU(N|N)$ theory.

Diagrams are constructed by Wick contracting (\ie creating propagators)
in expressions constructed out of supertraces (originating from the 
interactions). 
Ignoring the momentum dependence (since we are only interested here in the 
group theory factors) tree contributions formed from $\C$ propagators 
thus take the form:
\be
\label{ctree}
\str(X\C)\,\,\str(\C Y)\wick{19}{12}= \str(X Y)+\cdots.
\ee
Here $X$ and $Y$ are superfields or products of superfields.
We have used the freedom to cycle the two supertraces containing $\C$,
and combined them with \eq{cprop}. The ellipsis corresponds to the 
neglected terms in \eq{cprop}. 

These are closed into a $\C$ flavour loop by a further Wick contraction. 
The resulting terms have either already been shown to be finite (since 
they come from the symmetry broken 
part) or else without loss of generality the group theory part
takes the form:
\be
\label{cloop}
\str(\C X\C Y)\wick{13}{7}=\str X\, \str Y+\cdots,
\ee
where again we use \eq{cprop}, $X$ and $Y$ are (products of) remaining 
superfields, and the ellipsis is the neglected finite term generated 
by the second term in \eq{cprop}. 

The one-loop diagrams we are presently interested in
are thus given by a sum of contributions which are either already shown
to be finite or have the group theory structure $\str X\,\str Y$, 
where $X$ and $Y$ may contain in total up to
three gauge  fields. But such a term vanishes trivially, since either
$X$ or $Y$ must have one or less gauge field and thus yield
$\str\A=0$ or $\str\one=0$. 
Thus we see that $\C$ flavour One-loop Remainder Diagrams with $E_\A\le3$,
are finite as a consequence of power counting and these supertrace
identities.

\subsection{One-loop Remainder Diagrams with $\A$ propagators}

The analysis for $\A$-type loops proceeds similarly, however this time
with an extra twist because only 
$\A^A$ propagates (\cf sec. \ref{Alternatives}). 
For large momentum the $\A$ propagator behaves as
\be
\label{aprop}
<\!\A^A(p)\,\A^B(-p)\!>\,={c\over p^2} 
g^{AB} \left[ \delta_{\mu\nu}
+ {p_{\mu}p_{\nu} \over {p^2}} \left( {\ch\over\xi c}-1\right) 
\right]+O(p^{2\rt-4r-4}),
\ee
where we have used \eq{gmet} and \eq{igmet}, and
again we suppress the $p^2/\Lambda^2$ dependence of the cutoff functions.
The first term on the right hand side comes from \eq{YMact}
and \eq{gaugeact}, has the same form as in the unbroken theory and gives
the behaviour already accounted for in the power-counting analysis, and
the second term is the asymptotic behaviour coming from the 
regularised symmetry breaking mass term in \eq{brkCact}.

Once more we see by \eq{finalcons} that parts of the one-loop integral
involving symmetry breaking terms, are already finite:
to form the One-loop Remainder Diagrams we need to use pure $\A^i$ vertices
and these are either the unbroken ones from \eq{YMact}, giving the
index of divergence $2r+4-i$ as ascribed in the power counting analysis,
or again from the regularised mass term in \eq{brkCact} with index
$2\rt+2-i$. Thus if we use the symmetry breaking part of the propagator
in \eq{aprop} and/or the symmetry breaking vertices the 
degree of divergence of the resulting integral is bounded by 
$\DG\le D-E_\A-2(r-\rt+1)<0$. Once again these are the terms that
will be indicated only by the ellipsis.

From \eq{sow} and \eq{aprop}, the group theory part of 
tree contributions take the form:
\be
\label{atree}
\str(X\A)\,\,\str(\A Y)\wick{19}{12}= {1\over2}\,\str(X Y)+\cdots.
\ee
At first sight we should also add the terms
\be
\label{atree2}
-{1\over4N}\left(\tr X\str Y+\str X\tr Y\right),
\ee
coming from \eq{sow}.
These terms express the fact that only the parts which are 
both traceless and supertraceless, \viz
$X_R \equiv X-{\sig3\over2N}\str X-{1\over2N}\tr X$, couple to
$\A^A$. Indeed \eq{atree2} may be absorbed into \eq{atree} to
give ${1\over2}\str(X_R Y_R)+\cdots$. Nevertheless if \eq{atree2}
really remained, it would imply that 
the propagation of only\footnote{\ie without also $\A^0$ and/or $\A^{\sigma}$}
$\A^A$ is inconsistent since these terms arise in the unbroken theory
but $\tr X=\str\,\sig3 X$ (and {\it ditto} $Y$) breaks $SU(N|N)$.
Actually since all $\A$ interactions are through commutators,\footnote{in
the free-$\A^0$ representation or else extra interactions arise from the
*bracket, \cf sec. \ref{Alternatives}} by rearrangement we
can always express $X$ and $Y$ themselves as commutators, whence
$\str X$, $\str Y$ and \eq{atree2} actually vanishes.

By these arguments and \eq{atree}, such tree diagrams themselves are
supertraces of $\A$ times nested commutators (in the free-$\A^0$
representation) and thus for any given pair of (external) $\A$s 
in such a tree diagram, the group theory part may be expressed as a sum
of contributions of the form
\be
\label{atreetypes}
\str([\A,Z_1]Z_2[\A,Z_3]Z_4)\qquad{\rm and/or}\qquad\str(\A[\A,Z_1])
\ee
where the $Z_i$ are of course 
supermatrices. $Z_2$ and $Z_4$ could be $\one$
or non-trivial
(in which case in fact further commutators can be made).
Closing the resulting trees into an $\A$ flavour loop, by using 
\eq{aprop} and \eq{split},
the group theory part generically may be written:
\be
\label{aloop}
\str(\A X\A Y)\wick{13}{7}={1\over2}\,\str X\, \str Y+\cdots.
\ee
Here we have used the fact that only the block-diagonal, \ie super-group 
even, part of $X$ contributes (otherwise the Wick contraction connects 
$B$s to $A^i$s in \eq{defA} and trivially vanishes).
Again at first sight we ought to be including 
some unexpected $SU(N|N)$ breaking terms:
\be
\label{aloop2}
-{1\over4N}\tr(XY+YX),
\ee
coming from \eq{split},
however expanding the actual structures \eq{atreetypes} and summing over
\eq{aloop2} with the resulting $X$ and $Y$, one readily finds
that these terms vanish.

The net result is the same as the $\C$ loop: One-loop Remainder
Diagrams are given by a sum of contributions that either include
spontaneous symmetry breaking terms in which case they are finite, or
take the form of the unbroken theory in which case they are the product
of two supertraces, which vanishes for $E_\A\le3$.

\subsection{One-loop Remainder Diagrams with $\eta$ propagators}

The analysis of this case is virtually the same as for $\A$ above,
with the same conclusions,
unsurprisingly since $\eta$ is by BRST intimately related to gauge
transformations (see sec. \ref{Ward}).
The symmetry breaking mass term in \eq{ghostact} yields asymptotic
contributions whose $\DG$ is bounded above by $D-E_\A-2(\rh-\rt+1)$
and is thus already finite by \eq{finalcons}. 
Only the components $\eta^A$ and $\etab^B$ propagate
and thus individual tree contributions may result in terms of form
\eq{atree2}, but these vanish once we collect the interactions into
their commutator form. The same comments
apply to loops and \eq{aloop2}. 

\bigskip

To summarise, we have seen that whatever flavour is involved in the
loop, those parts of the One-loop Remainder Diagrams associated with the 
spontaneous breaking of $SU(N|N)$ are finite by \eq{finalcons}.
(Note that this includes the second term in
the gauge fixing function \eq{gafif}, and
the corresponding terms in \eq{gaugeact} and \eq{ghostact}.)
Although we have concentrated on the One-loop Remainder Diagrams, it
is clear from the preceeding analysis that all one-loop
unbroken $SU(N|N)$ contributions, 
have double supertrace form
$\str X\,\str Y$, where $X$ and $Y$ are (products) of the external $\A$s,
and thus vanish for $E_\A\le3$. Consequently One-loop Remainder Diagrams
with three or less external gauge fields are finite by power counting
and the supertrace mechanism.

\subsection{Large $N$ limit}
\label{Large}

It is appropriate to note here that in the large $N$ limit all these
unbroken $SU(N|N)$ contributions vanish. The large $N$ limit for
Yang-Mills is achieved
by rescaling $g^2$ to $g^2/N$. As a loop-counting parameter this is
balanced by those terms which contribute an extra $\tr\one=N$ at each
new loop order resulting in a non-trivial limit \cite{thooftln}.
In our case, these double (super)trace terms 
are down by a factor $1/N$, unless one of them is empty but then the
result vanishes by $\str\one=0$. It follows in particular
that all the One-loop Remainder Diagrams are thus finite in the large
$N$ limit, and thus we have proved that in the $N=\infty$ limit, the
theory is finite in all dimensions $D$.

In fact in this way there are no factors of $N$ coming
from loops to balance the rescaled $g^2/N$, since $\tr\one=N$
has been replaced by $\str\one=0$. Thus as a consequence of the supertrace
mechanism, in the large $N$ limit the symmetric phase $SU(N|N)$ theory has 
no quantum corrections at all \cite{manerg}\cite{erg1}\cite{erg2}.


\section{Ward identities}
\label{Ward}

At finite $N$, we have not ruled out the possibility of divergent one-loop 
contributions with $3<E_\A\le D$ external $\A$s (and no external $\C$s
or ghosts) originating from the unbroken theory,
however we have yet to use the constraints of gauge invariance, 
which limit the possible divergences (just as they do for quantum 
electrodynamics and ordinary Yang-Mills, \eg in the finiteness of
the four-photon vertex, and the Slavnov-Taylor identities respectively).
We will show that these remaining diagrams are in fact finite in all
dimensions $D<8$. Since all other corrections have already been shown
to be superficially finite in any dimension,\footnote{for suitable
choice of ranks $r,\rt,\rh$, \cf \eq{finalcons}} the finiteness to
all orders in perturbation theory of the full theory is then proved
for all dimensions $D<8$.

Since the contributions in question arise from the unbroken parts of
the Lagrangian only [including only
those generated by the first term of 
\eq{gafif}], we may as well work with unbroken Ward identities:
this simplifies the arguments, and the broken Ward identities only
introduce further terms which as we have already seen in sec. 
\ref{Supertrace}, are finite, given \eq{finalcons}, already by power 
counting. 

Working from now on in this section with the unbroken phase, we remind
that the one-loop two and three-point pure $\A$ vertices
actually vanish by the supertrace mechanism (\cf sec. \ref{Supertrace}). 
In $D=4$ dimensions for example, 
considerations of renormalizability would make it
very surprising if the one-loop four-point vertex then turned out to
diverge! 

Indeed, gauge transformations equate any longitudinal part of the 
four-point vertex to the sum of three-point vertices.
(See for example the discussion of such identities in refs. 
\cite{erg1,erg2}. The explicit equation is that given by just the first 
three terms in \eq{thebiz}.)
Since the three-point
vertex vanishes, the four-point vertex must be transverse on all four
legs. This is only possible if the diagram works out to have a tensor
structure involving at least four external momenta.\footnote{An 
$\A_\mu(p)$ external line
must have tensor structure $\delta_{\mu\alpha}p_\beta-\delta_{\mu\beta}
p_\alpha$ where $\alpha$ and $\beta$ are other external indices or
contract into indices in the rest
of the diagram. This is the structure of $\F_{\alpha\beta}$.}
This means that there are four less powers of loop momentum available, so
the superficial degree of divergence of the one-loop four-point vertex drops 
from $\DG=D-4$ to $\DG=D-8$. In other words the one-loop four-point 
pure $\A$ vertex is finite in all dimensions $D<8$.

Proceeding in this way, we can show the finiteness of all the remaining
vertices. Thus any longitudinal part of the five-point pure $\A$ vertex is
equal to the sum of four-point vertices, and thus is finite for all $D<8$. 
All that remains is a totally transverse part which by the arguments above
actually has $\DG=D-5-5$,
and thus is finite for all $D<10$. By iteration, we see that
for all dimensions $D<8$ the remaining $3<E_\A\le D$ One-loop Remainder
diagrams are finite as a consequence of power counting, the supertrace
mechanism and gauge invariance.

However, the above arguments are only strictly valid in a scheme 
such as developed in ref. \cite{erg2}, in which
manifest gauge invariance is maintained at all stages. In order for
the arguments to be rigorous in this more traditional 
gauge fixed context, we must demonstrate the existence
of the corresponding BRST invariance and develop the appropriate 
Lee-Zinn-Justin identities.

One can readily check that with the multiple grading assignments 
of \sec{Alternatives}, the usual BRST algebra:
\bea
\label{brst}
\delta \A_\mu &=&\epsilon\Lambda^{D/2-2}[\nabla_\mu,\eta]\nonumber\\
\delta\C &=& -ig\epsilon[\C,\eta] \nonumber\\
\delta\eta &=& ig\epsilon\eta^2 \nonumber\\
\delta\etab &=& \epsilon\xi\Lambda^{D/2-2} \ctil {F_{symm}}
\eea
is an invariance of the na\"\i ve functional measure and the
unbroken action $S_{YM}+S_C+S_{Gauge}+S_{Ghost}$,
where $S_{YM}$ is given in \eq{YMact}, $S_C$ is the unbroken version
\eq{unbCact}, $S_{Gauge}$ utilises only
$F_{symm}=\partial_\mu\A_\mu$ which is the gauge fixing
function \eq{gafif} but discarding the second part referring to breaking,
and similarly the ghost action refers only to the first
part of \eq{ghostact}. ($\epsilon$ has been defined dimensionless.
Of course it is straightforward to write the
BRST algebra and so forth for the broken Ward identities and/or more
general gauge fixing functions, but not helpful in the present context.)

The derivation of the Lee-Zinn-Justin identities proceeds in standard
fashion \cite{zinn}.\footnote{except for care with sources (fields) that do
not couple to $\sig3$ ($\one$), and commutation} As usual
we add to the action source terms for the fields and the non-linear BRST 
transformations, however it is helpful to express them as supermatrices and
contract using the supertrace:
\be
\label{sources}
S_{Sources} 
=-\,\str\!\!\int\!\! d^{D}\!x\,\left(\J_\mu\A_\mu+\J\C+\zetab\eta+
\etab\zeta+\Lambda^{{D\over2}-2}
\K_\mu\nabla_\mu\!\cdot\!\eta-ig\H[\C,\eta]+ig\L\eta^2\right).
\ee
The sources thus live in the dual space 
as determined by the Killing metric \eq{killing} and \eq{massacre}. 
Therefore
\be
\J=\left( \begin{array}{cc} J^1 & K \\
		       \bar{K} & J^2 
	       \end{array}
       \right),                 
\ee
is an unconstrained superfield, but $\J_\mu$ (distinguished from $\J$ by 
the Lorentz index) expands only over $T_A$ and $\sig3$ (or just over $T_A$ 
for the *bracket formalism of sec. \ref{Alternatives}):
\be
\J_\mu=2\J^A_\mu T_A+{1\over2N}\J^{\sigma}_\mu\sig3\qquad{\rm so}\qquad
\str\,\J_\mu\A_\mu=\J^A_\mu\A_{A\,\mu}+\J^{\sigma}_\mu\A^0_{\mu},
\ee
the same constraints applying for all the other sources: $\zeta$, $\zetab$,
$\K$, $\H$ and $\L$. We define the
functional differentials of source or field so as to extract the
conjugate from under the supertrace thus \cite{erg2} 
\be
\label{djdef}
{\delta \over {\delta\J}} := {
\left(\!{\begin{array}{cc} {\delta / {\delta J^1}} & - {\delta /
{\delta \bar{K}}} \\ {\delta / {\delta K}} & - {\delta
/ {\delta J^2}} \end{array}} \!\!\right)},
\ee
so
\be
{\delta \over {\delta\J}}\; \str\!\!\int\!\! d^{D}\!x\,\J\C =\C,
\ee
with a similar definition for $\delta/\delta\C$, whilst 
\be
{\delta\over\delta \J_\mu}:=T_A {\delta\over\delta\J_{A\,\mu}}
+\one{\delta\over\delta \J^\sigma_\mu}
\ee
has the same effect on the $\J_\mu\A_\mu$ term ($\delta/\delta\J_{A\,\mu}
=g^{AB}\delta/\delta\J^B_\mu$), the other source and field differentials
being defined similarly, for example 
\be
{\delta\over\delta\A_\mu}:=
2T_A{\delta\over\delta\A_{A\,\mu}}+{\sig3\over2N}
{\delta\over\delta\A^0_\mu}.
\ee
First order variation over sources (the chain rule) is 
then simply given by
\be
\label{chainrule}
\str\!\!\int\!\! d^{D}\!x\left(\delta\J_\mu {\delta\over\delta\J_\mu}
+\delta\J {\delta\over\delta\J}+\delta\zeta {\delta\over\delta\zeta}
+\delta\zetab {\delta\over\delta\zetab}\right),
\ee
with of course a similar expression for the fields.

Under the BRST transformations \eq{brst}, the generator of connected 
diagrams $W=\ln {\cal Z}$ then satisfies
\be
\xi\Lambda^{D/2-2}
\zeta\cdot\ctil\cdot\partial_\mu{\delta W\over\delta\J_\mu}+
\str\!\!\int\!\! d^{D}\!x\left(
\J_\mu{\delta W\over\delta\K_\mu}+\J{\delta W\over\delta\H}-\zetab
{\delta W\over\delta\L}\right)=0.
\ee
Legendre transforming to the generator of 1PI diagrams:
\be
\Gamma
+\xi\,\partial_\mu\A_\mu\cdot\chat\!\cdot\partial_\nu\A_\nu
=-W+\str\!\!\int\!\! d^{D}\!x\left(\J_\mu\A_\mu+\J\C+\zetab\eta+
\etab\zeta\right),
\ee
where $\A_\mu$, $\C$ and $\eta$ are now classical fields. We have
extracted the gauge fixing term, so that on using the antighost 
Dyson-Schwinger equation\footnote{the restriction to the dual
of $S_\alpha$ arising from \eq{gensum}, \ie 
$\delta\etab=\delta\etab^\alpha S_\alpha$.}
\be
\str\ S_\alpha\!\left({\delta\Gamma\over\delta\etab}\Lambda^{D/2-2}
-2\chat\ct\,\partial_\mu{\delta\Gamma\over\delta\K_\mu}\right)=0,
\ee
we obtain the simplified Lee-Zinn-Justin identities:
\be
\label{lzjid}
\str\!\!\int\!\! d^{D}\!x\left(
{\delta\Gamma\over\delta\A_\mu}{\delta\Gamma\over\delta\K_\mu}+
{\delta\Gamma\over\delta\C}{\delta\Gamma\over\delta\H}+
{\delta\Gamma\over\delta\eta}{\delta\Gamma\over\delta\L}\right)=0.
\ee

Before we can use these to establish finiteness of the remaining One-loop
Remainder Diagrams, we have to investigate the finiteness
of the new diagrams (in the full broken theory) 
involving interactions introduced by
the BRST sources $\K_\mu$, $\H$ and $\L$ in \eq{sources}.
It turns out that
since these interactions do not involve the higher derivatives,
all such diagrams are superficially finite by power counting. 
It is straightforward to adapt the arguments of sec. \ref{Counting}
to prove this. We sketch the alterations.
We note that \eq{sdd1im} is unchanged, however eqns \eq{earel}--\eq{eerel}
pick up corrections from the BRST source
vertices. The ghost equation in the desired form \eq{alteerel}
is however unchanged, as can be most simply understood again by deriving the 
equation directly as a count over external 
antighosts.\footnote{Alternatively incorporate $\K$, $\H$ and $\L$ in the 
ghost-number conservation equation and note that these sources always appear
as many times as the corresponding vertices.} The result is that 
(for 1PI diagrams) $\DG$ in form \eq{sdd2} picks up the new terms
\be
\label{sdd2khl}
-(2r+3)E_\K-(r+\rt+3)E_\H-(2r+4)E_\L,
\ee
whilst Proposition 1 holds unchanged. We obtain the same sufficient
conditions \eq{conditions} and \eq{1looprels} since corrections \eq{sdd2khl} 
are negative with these, which as before regulate
all but a small set of diagrams. But before refinement, this latter set
now contains diagrams with external BRST sources, since the constraint
that there be no external antighosts no longer implies that there are no
external ghosts.  However, by conditions \eq{1looprels}, \eq{sdd2} and
\eq{sdd2khl}, all diagrams containing
$\K$, $\H$ or $\L$ already result in $\dgaone<0$. Thus under the
earlier necessary and sufficient conditions on $r$, $\rt$ and $\rh$, \viz \eq{1looprels2} 
equivalently \eq{finalcons}, all but the same set of One-loop Remainder Diagrams 
have $\DG<0$, and in particular all the diagrams involving BRST sources are 
superficially finite in any dimension $D$.

Finally, working to one-loop and writing $\Gamma$ in terms of its classical
and one-loop parts, $\Gamma=\Gamma^0+\hbar\Gamma^1$, keeping the
$O(\hbar)$ terms of \eq{lzjid}, and extracting from that those terms with 
one $\eta$ and otherwise only $\A$s, we obtain, up to unimportant
corrections which are finite in all dimensions, precisely the Ward 
identities used at the beginning of this section to prove finiteness in all
dimensions $D<8$. 

To be explicit, we 
write in the unbroken theory the one-loop pure $\A$ vertices as
\bea
\label{exone}
{1\over2!}\sum_{m,n=2}^\infty{1\over nm}\int\!\!
d^D\!x_1\cdots d^D\!x_n\,d^D\!y_1\cdots&&\!\!\!d^D\!y_m\,
\Gamma^1_{\mu_1\cdots\mu_n,\nu_1\cdots\nu_m}
(x_1,\cdots,x_n;y_1,\cdots,y_m)\\
&&\str\, \A_{\mu_1}(x_1)\cdots \A_{\mu_n}(x_n)\
\str\, \A_{\nu_1}(y_1)\cdots \A_{\nu_m}(y_m),\nonumber
\eea
using the conclusions of sec. \ref{Supertrace}. The supertrace
structure implies that the vertices are cyclic
on the $x_i^{\mu_i}$ arguments and $y_j^{\nu_j}$ arguments separately, 
and symmetric under exchanging the two sets of arguments (see also
\cite{erg1}\cite{erg2}), and that the vertices $\Gamma^1$ may be defined to
vanish identically for $n$ or $m$ less than 2. The $O(\hbar)$ terms in 
\eq{lzjid} with one $\eta$ and otherwise only $\A$s, only come from the 
terms 
\be
\label{ward2}
\str\!\!\int\!\! d^{D}\!x\left(
{\delta\Gamma^1\over\delta\A_\mu}{\delta\Gamma^0\over\delta\K_\mu}+
{\delta\Gamma^0\over\delta\A_\mu}{\delta\Gamma^1\over\delta\K_\mu}\right),
\ee
and thus
\bea
\label{thebiz}
p_1^{\mu_1}\Gamma^1_{\mu_1\cdots\mu_n,\nu_1\cdots\nu_m}
(p_1,\cdots,p_n;q_1,\cdots,q_m)= 
\Gamma^1_{\mu_2\cdots\mu_n,\nu_1\cdots\nu_m}
(p_1\!+\!p_2,p_3,\cdots,p_n;q_1,\cdots,q_m) \nonumber\\
 -\Gamma^1_{\mu_2\cdots\mu_n,\nu_1\cdots\nu_m}
(p_2,\cdots,p_{n-1},p_n\!+\!p_1;q_1,\cdots,q_m)\ +\ {\rm finite},
\hspace{1cm}
\eea
Ward identities for the other arguments follow from cyclicity and 
exchange symmetry. The explicit terms are precisely those of the 
Ward identities \cite{erg1,erg2} we already referred to and used at the 
beginning of this section and follow from the 
first term in \eq{ward2}. The only change is the addition of the term
``finite'' which comes from the second term in \eq{ward2}; this is finite
by \eq{sdd2khl} and the arguments below, 
as a consequence of the fact that only the terms in $\Gamma^1$ containing at
least one $\K$ contribute.
We thus see that any longitudinal part of the four ($n=m=2$) point
vertex is finite in any dimension, leaving only 
the possibility of divergences which are totally transverse.
However, by power counting such a term is finite in all dimensions 
$D<8$ as shown 
at the beginning
of this section.\footnote{Out of interest we note from \eq{exone}
that it is the coefficient of $(\str\,\F^2_{\mu\nu})^2$ that diverges in
$D=8$.}
The rest of the arguments from the beginning of this
section follow through similarly.

\section{Unitarity}
\label{Unitarity}

The $A^2_\mu$ and $C^2$ fields of \eq{defA} and \eq{defC} have wrong
sign actions as a consequence of the supertrace. Na\"\i vely these
functional integrals in the partition function do not make sense,
however the correct prescription is to analytically continue these
functional integrals whilst respecting $SU(N|N)$.  Equivalently we may
define the system through exact renormalization group methods
\cite{erg1}\cite{erg2}\cite{catchall} or operator methods, neither of
which suffer difficulties of definition. Actually, there is a choice of
Fock vacuum (\viz annihilators) violating
$SU(N|N)$, and resulting in an unbounded Hamiltonian. (This in turn
would signal an unstable theory.) But covariant quantization leads to
a bounded Hamiltonian.  The problems of wrong sign action then show up
in the appearance of negative norm states, which are thus unphysical
and lead to a non-unitary S matrix. Below, we demonstrate these points
on a simple quantum mechanics example. In
the continuum limit $\Lambda\to\infty$, all fields apart from the $A^i_\mu$ 
become infinitely massive and, as we will see in the second subsection, 
for $N=\infty$ and any dimension $D$, or for finite $N$ but
providing $D\le4$ dimensions, the unphysical $A^2_\mu$ field completely 
decouples from the physical $A^1_\mu$. In this way, a unitary $SU(N)$ 
Yang-Mills theory is recovered in the limit $\Lambda\to\infty$.

\subsection{$U(1|1)$ quantum mechanics example and negative norms}

Defining the Hermitian super-position $\X$ as  
\be
\X = \left( \begin{array}{cc} x^1 & \altpsi \\
		       \bar{\altpsi} & x^2 
	       \end{array}
       \right),                 
\ee
we consider the Minkowski type Lagrangian of a simple harmonic
potential:  $L = {1\over 2}\str\Xdot^2 - {1\over 2}\str\X^2$.
Classically this Lagrangian is invariant under $SU(1|1)$
transformations $\del\X=i\,[\omega,\X]$, however we buy for free
invariance under the full $U(1|1)$. By Noether's theorem these are
generated by the triplet of charges (\aka angular momenta)
\be
\label{defQ}
\Q=i\,[\X,{\dot\X}],
\ee
through the Poisson bracket with $\str\,\omega\Q$.
Note that the charge for $\omega\sim\one$, vanishes, reflecting its
trivial action on $\X$. Defining a
super-covariant derivative as in \eq{djdef}, the supermomentum is
\be
\label{bigmom}
{\cal P} := {\partial  \over {\partial \dot{\X}}} L = \dot{\X},
\ee
and differs by some convenient signs from the usual definitions:
\be
\label{usualp}
p_i={\partial L\over\partial {\dot x}^i},\qquad 
p_\altpsi={\partial L\over\partial {\dot\altpsi}},\qquad 
p_{\bar\altpsi}={\partial L\over\partial {\dot{\bar\altpsi}}}.
\ee
The Hamiltonian is then given as
\be
H=\str\,{\cal P}{\dot\X}-L,
\ee
and quantization is via the graded commutator:
\be
\left[(\X)^a_{\gap b}\, ,({\cal P})^c_{\gap d}\right]_\pm
= i (\sig3)^a_{\gap d}\,\del^c_{\gap b}.
\ee
This is the form that respects $U(1|1)$, as can most easily be seen by
writing it contracted with arbitrary constant supermatrices $U$ and $V$:
\be
[\str U\X,\str V\P]=i\,\str\, UV,
\ee
and actually corresponds to the usual relations for the usual definitions
of momenta \eq{usualp}.
However, as often happens, we have to be careful with operator ordering
since the na\"\i ve ordering implied by \eq{defQ}, on quantization no
longer leaves $\Q$ supertraceless. We can cure this by subtracting the
supertrace which as we will see, for a sensibly defined vacuum,
corresponds to normal ordering:
\be
\label{defQQ}
\Q=i\,[\X,\P]-{i\over2}\sig3\,\str[\X,\P]=i\,[\X,\P]+2\sig3.
\ee
The definition of the vacuum follows from the choice of annihilation
and creation operators
\be
\label{creation}
A=(\X+i\P)/\sqrt2,\qquad A^\dagger=(\X-i\P)/\sqrt2,
\ee
with normalised vacuum such that
$A\vac=0$. The $A$s have the now expected graded commutation relations,
\be
\label{AAd}
\left[(A)^a_{\gap b}\, ,(A^{\dagger})^c_{\gap d}\right]_\pm =  
(\sig3)^a_{\gap d}\,\del^c_{\gap b}.
\ee
It is straightforward to check that the vacuum respects $U(1|1)$:
$\Q\vac=0$. As advertised, the supercharges \eq{defQQ} alternatively
may be written $\Q=\,:[A^\dagger,A]:\,$.

Writing \eq{creation} in terms of components
and using the usual definitions of momenta \eq{usualp}, $x^1$ has
the usual form for an annihilation operator, $a^1=(x^1+ip^1)/\sqrt2$, but
$x^2$ has an annihilation operator containing
the wrong sign: $a^2=(x^2-ip^2)/\sqrt2$. These components
therefore have the wrong sign commutation relations $[a^2,{a^2}^\dagger]
=-1$, as is easily seen from \eq{AAd}. This is just what is needed to
compensate the wrong sign for ${a^2}^\dagger a^2$ in the Hamiltonian, 
$H=\str\,A^\dagger A+2$, 
which is thus bounded below. However, negative norms appear in
the `2' sector:
\be
|n_2\rangle = {1 \over\sqrt{n_2 !}}(a^{2\dagger})^{n_2} \vac,\qquad 
\langle n_2|n_2\rangle=(-1)^{n_2}.
\ee
It is straightforward to verify that any attempt to repair this
by keeping $a^1$ as it is, but changing the sign in $a^2$, results in
an unbounded Hamiltonian and a vacuum that violates both
$U(1|1)$ and $SU(1|1)$: $\Q\vac\ne0$.

(Let us note that this situation is very similar
to the Gupta-Bleuler quantisation procedure~\cite{itz} which has to
deal with a wrong sign action for time-like photons. The same choice of
vacua exists, but Lorentz covariant quantization picks out the one with
negative norm states. In our case however, we have no equivalent
Gupta-Bleuler condition for excluding such unphysical states. Instead,
the unphysical sector decouples, as we described in the introduction
to this section, and now show.)


\subsection{Recovery of unitarity in the $A^1$ sector}
\label{Recovery}

We have seen that our covariant higher derivative spontaneously broken 
$SU(N|N)$ theory is finite in all dimensions $D<8$. However, this is not
enough to show that it acts as a regulator for $SU(N)$ Yang-Mills theory.
For this to be the case, we must show that for renormalized variables in the
continuum limit $\Lambda\to\infty$, $SU(N)$ Yang-Mills
theory is recovered. 

All fields but the $SU(N)\times SU(N)$ 
gauge fields $A^i_\mu$ (and when gauge fixed their respective 
ghosts $\eta^i$), become infinitely massive in this limit and thus drop
out of the spectrum. The issue then is to show that there are no remaining
effective interactions between these two gauge fields: the wrong sign $A^2$
sector can then just be ignored. From above, it is also necessary to
establish that unitarity is recovered in this limit. But this is the
same question, since a non-unitary amplitude in the $A^1$
sector can only arise in the $\Lambda\to\infty$ limit, from contributions
with internal $A^2$s. Cutkosky cutting  such an amplitude
must then result in a non-vanishing amplitude connecting $A^1$s and
$A^2$s \cite{itz}. Therefore providing that we can establish that there are 
no such effective interactions between $A^1$ and $A^2$, we can safely ignore
the sick $A^2$ sector and recover a unitary continuum limit for the
$SU(N)$ Yang-Mills theory.

Actually, in the large $N$ limit, there is nothing further to do: 
since only single trace interactions survive \cite{thooftln} (see also 
\cite{manerg}\cite{erg1}\cite{erg2} and subsec. \ref{Large}) and any 
interaction between $A^1$ and $A^2$ (or their ghosts) requires two
traces, one for each $SU(N)$, the separation of the two sectors is
automatic. As shown in \sec{Supertrace}, in this limit the theory is
finite in all dimensions. Therefore
in the $N=\infty$ limit, our regularisation works for
$SU(N)$ Yang-Mills in any dimension $D$.


For finite $N$, we appeal to the Appelquist-Carazzone decoupling
theorem \cite{apple} to show that the regularisation works for any dimension
$D\le4$. The theorem states that for a renormalizable theory,
as the mass scale of the heavy sector tends to infinity, the (bare)
effective Lagrangian is given by a renormalizable Lagrangian for the
light fields with irrelevant corrections vanishing by inverse
powers of the heavy scale. This scale is identified with the overall cutoff
for the effective theory. This theorem for example justifies the assumption
that a spontaneously broken GUT (Grand Unified Theory) is equivalent to the
Standard Model $SU(3)\times SU(2)\times U(1)$ at energies where the
GUT scale can be neglected, and our case is closely analogous with the 
$SU(N|N)$ theory playing the r\^ole of the GUT. Just as there would
be no interactions between the $SU(3)$, $SU(2)$ and $U(1)$ gauge fields of
the Standard Model if it were not for the matter fields, as we confirm
below there are no interactions between the two $SU(N)$ gauge fields
in our effective theory. 

Note that the Appelquist-Carazzone theorem applies only to an initially
renormalizable theory. The spontaneously broken $SU(N|N)$ theory without
the higher derivatives is renormalizable in $D\le4$ dimensions
(because the standard analysis does not care that we are dealing with a 
supergroup). The higher derivatives are a regularisation 
for this theory. In point of fact, our situation is simpler than the
cases considered for the original proofs where the essential difficulty
arises from the exchange of limits of the heavy scale tending to infinity
and the overall cutoff tending to infinity \cite{apple}. 
In our case the two scales are identified and by construction
in sec. \ref{Spontaneously}, the only scale in the theory is $\Lambda$.

Thus in $D\le4$ dimensions, the
effective $SU(N)\times SU(N)$ theory can be described by an effective
bare Lagrangian containing only these fields, their own 
Yang-Mills couplings $g_i$ (no longer equal to $g$) and further interactions
weighted by appropriate powers of $\Lambda$ as determined by dimensions. 
All of these other interactions are however irrelevant and vanish in the
limit $\Lambda\to\infty$. In particular, the lowest dimension
interaction between the two fields comes from a group theory structure
$\tr\, A^1_\mu A^1_\nu \ \tr\, A^2_\lambda A^2_\sigma$,
(with Lorentz indices contracted in some way). Since such
an interaction must also be gauge invariant under $SU(N)\times SU(N)$, 
the minimal dimension bare interaction actually takes the form
\be
\Lambda^{-D}\tr\, F^1_{\mu\alpha} F^1_{\nu\beta}\ 
\tr\, F^2_{\lambda\gamma} F^2_{\sigma\delta},
\ee
which is irrelevant in any dimension. (Here $F^i_{\alpha\beta}$ is the 
field strength for $A^i_\mu$, the Lorentz indices are again contracted in 
some fashion, and the $\Lambda$ dependence is displayed up to $\ln\Lambda$ 
multiplicative corrections.) 

As in the initial arguments of sec. \ref{Ward}, we 
have assumed gauge invariance, whereas we have been working within a
traditional gauge fixed approach. The extra details coming from ghosts
and BRST do not change the conclusions and have already been
treated in earlier work on the decoupling theorem \cite{kazama}.

Finally, we have seen that at finite $N$, $D\le4$ is a sufficient condition
for decoupling. It is also necessary, since in
$D>4$ dimensions the couplings $g_i$ are
non-renormalizable, and thus clearly  
all higher order interactions will be unsuppressed.

\section{Convergence}
\label{Preregularisation}

In this section we explain precisely why a preregularisation
is needed, why dimensional regularisation is sufficient - even 
non-perturbatively, and why at $N=\infty$ or when $D<4$,
the system is well defined even
without preregularisation. To help in explaining this,
we first discuss preregularisation in PV systems more generally. 

Any PV regularisation scheme arrives at finite 
results by the addition of separately divergent quantities. Thus care must be 
taken to define these conditionally convergent integrals unambiguously,
and importantly, in a way which is consistent with gauge invariance.
(An example of this problem was given in a lecture on our
earlier $N=\infty$ version \cite{manerg}, where a typical integral that 
formally vanishes after a shift in momentum of one
contribution, was shown to yield a positive answer by equally formal 
manipulations.) 

Traditional PV regularisations
solve the problem by so-called momentum-routing \cite{itz}: diagrams can
be collected together so that any internal
propagator is accompanied by PV propagators 
{\sl with precisely the same momentum}, for example:
\be
{1\over p^2+m^2}-{1\over p^2+M^2}.
\ee
Although separately the contributions are divergent, the result is
convergent, and thus only conditionally so. Now by algebraic addition,
all such contributions are replaced by terms with  sufficiently fast decay
that the integrals are absolutely convergent and thus unambiguously defined.
In our example the large momentum behaviour $1/p^2$ of each individual
piece is replaced by one piece with large momentum behaviour $1/p^4$:
\be
{M^2-m^2\over (p^2+m^2)(p^2+M^2)}.
\ee

Bakeyev and Slavnov solved the problem for their gauge invariant PV system 
by the addition of further PV fields in such a way that quantum corrections 
do indeed result from cancellations between
diagrams with the same structure, allowing again an unambiguous (and 
eventually gauge invariant) definition by labelling
the momenta of the corresponding diagrams in the same way \cite{pvs}.
However, the gauge invariant PV systems of refs. \cite{pvo,pva,pvp}, 
\cite{manerg}--\cite{romerg} and this paper, cannot use such a technique 
directly, because the regulating diagrams do not all have the same structure
as the original Feynman graphs.

Actually, any precise Poincar\'e invariant 
rule for evaluating the momentum integrals is acceptable 
providing it yields the correct finite answer on absolutely convergent integrals 
(of course), and is invariant under shifts in the loop momenta. As emphasised 
in ref. \cite{manerg}, invariance under such shifts is the crucial property 
that ensures gauge invariance. Equivalently, this requires ensuring that all
the surface terms at infinite momentum that result from such shifts,
are discarded, even if they 
are non-vanishing. Providing that it is not necessary to keep a fixed dimension 
at intermediate stages
(for example to study chiral or topological phenomena), the problem can be
solved very straightforwardly by computing all integrals in general dimension
$D$, where such surface terms can easily be discarded,\footnote{typically
automatically} and only sending $D$ to the correct dimension at the end of the 
calculation \cite{manerg}--\cite{romerg}.

Let us emphasise that our computation
of the correct one-loop $\beta$ function \cite{manerg}--\cite{romerg}, 
which used essentially this $SU(N|N)$
regularisation and evaluated the momentum integrals in this way, 
already demonstrates how these ideas work in a practical
example (even if this computation was also done without fixing the gauge).

Note that the procedure amounts to using dimensional regularisation as a 
{\sl preregular\-isation}, however it is {\sl not} used
to identify divergences. Nor do we employ subtraction schemes relying on
dimensional regularisation (like the $MS$ or $\overline{MS}$ renormalization
scheme). The usual difficulties
in applying dimensional regularisation non-perturbatively (for example
in the large $N$ approximation \cite{zinn,ln}, where divergences
no longer appear as poles in $4-D$) therefore do not apply here. 
This is in contrast to the momentum-routing solution to PV regularisation 
outlined above, which
does not make sense beyond Feynman diagrams, \ie perturbation theory.

Dimensional preregularisation employed in the above way is our method
of choice when $N$ is finite and $D\ge4$, however at $N=\infty$ or if 
the dimension $D<4$, no preregularisation is in fact required (and thus we can
work in fixed dimension): the structure 
of the $SU(N|N)$ supergroup supplies an unambiguous recipe for evaluating
all the conditionally convergent integrals, by reorganising them into
absolutely convergent integrals. To see this, recall the crucial steps of the 
proof of finiteness: in \sec{Counting} we isolated the One-loop
Remainder Diagrams, which are the only (sub)diagrams not already
superficially finite by power counting (after suitable choices are made for
$r$, $\rt$ and $\rh$). Given finiteness overall, the One-loop
Remainder Diagrams are then by definition those pieces that are only
conditionally convergent. All the other momentum
integrals are already absolutely convergent. 

In \sec{Supertrace}, we showed
that in these remaining (sub)diagrams, all the bits associated with 
spontaneous symmetry breaking are also absolutely convergent. (Thus we expand
these one-loop integrals by writing all propagators and vertices as those
of the symmetric phase plus differences. The analysis of \sec{Supertrace}
shows that any resulting part including any one of these differences, is 
already finite by power counting, and thus absolutely convergent.)
We saw that $SU(N|N)$ group theory alone is enough
to ensure the {\sl exact cancellation} of the unbroken bits with $E_A\le3$
external $\A$ legs (and at $N=\infty$, any number of $\A$ legs). 
Note that these cancellations take place algebraically even
before we consider performing the momentum integrals. Thus by performing
the calculations in a way that respects global $SU(N|N)$ invariance 
(for example as we indicated in \sec{Supertrace}) we define the conditionally
convergent one-loop (sub)diagrams with $E_A\le3$ by reducing them to
absolutely convergent contributions (containing at least one term
associated with the spontaneous symmetry breaking). Since this covers
all the One-loop Remainder Diagrams in dimension $D<4$, we have demonstrated 
absolute convergence for the complete theory in this case.
At $N=\infty$, this
algebraic reduction takes place for all $E_\A$ and thus for all the 
One-loop Remainder Diagrams in any dimension $D$.
As claimed, preregularisation is thus
unnecessary for the cases $D<4$ and/or $N=\infty$.

The reader may wonder what is the obstruction to using this proof 
to define, without preregularisation, all the conditionally convergent cases 
when $N$ is finite and $4\le D<8$. 
To do this, we would have to use the local $SU(N|N)$ Ward identities of 
\sec{Ward} to {\sl define} by iteration
the  remaining only conditionally convergent contributions, \viz the 
symmetric phase parts of the One-loop Remainder Diagrams with $4\le E_\A\le D$,
starting with the $E_\A=4$ point vertex.
Any longitudinal part of this vertex can be cast 
by the Ward identities \eq{thebiz} into absolutely convergent
integrals. As we noted in \sec{Ward}, this means that any potential remaining 
divergence would have to have a tensor structure which is 
totally transverse and thus be a polynomial of minimum degree four in the 
external momenta. By power counting, this reduces the degree of divergence 
$\DG=D-4$, by four, and thus proves convergence in all dimensions $D<8$.
This last step in the argument however,
relies upon the fact that all the {\sl integrands} are {\sl local in the 
external momenta} \ie contain no explicit $1/p^2$ in external momenta $p$
which could instead compensate for the polynomial appearing
in the tensor structure. This is where the problem lies in extending this 
to prove absolute convergence: 
the full expression for the totally transverse remainder,\footnote{as opposed 
to  any potential divergence} defined either 
by subtracting the absolutely convergent longitudinal parts using 
longitudinal projectors ($p^\mu p^\nu/p^2$), or directly by multiplying all 
external legs with transverse projectors ($\delta_{\mu\nu}-p^\mu p^\nu/p^2$), 
is not local in the external momenta.

\section{Conclusions}
\label{Conclusions}

We constructed a regularised extension of pure $SU(N)$ Yang-Mills 
theory based upon a $SU(N|N)$ gauge theory with a 
Higgs super-field. Such a regularisation scheme meets all the 
requirements listed in the Introduction and is suitable for 
application within the gauge invariant ERG approach. To the best 
of our knowledge it is the first example of a regularisation which 
satisfies these conditions. 

In this section, we first recall from \sec{Alternatives} the main issues 
encountered in building the spontaneously broken $SU(N|N)$ gauge theory. 
Then we summarise
the steps of the proof that, together with covariant higher derivatives,
this provides a regularisation for $SU(N)$ Yang-Mills. Finally we draw
our conclusions noting in particular further properties and extensions.

We first noted that $U(N|N)$ is not reducible to a
product $SU(N|N)\times U(1)$, removing the usual {\it a priori}
reason for not considering such a group. The enlarged
group $U(N|N)$ requires an extra gauge field $\A^\sigma$.
However if we build a gauge theory on
$U(N|N)$, then $\A^0$ acts as a Lagrange multiplier
constraining $\A^\sigma$ to be pure gauge. Effectively, $U(N|N)$
contracts dynamically to $SU(N|N)$. Working directly
with $SU(N|N)$, the gauge field $\A^0$, being associated with the $\one$
generator, appears nowhere in the Lagrangian since the theory contains only
adjoint fields. The partition function
thus contains a free functional integral over $\A^0$ (\ie without
even a Gaussian weight). We cannot simply delete the $\A^0$ field
however because it is needed to absorb gauge transformations in the $\one$
direction. This is the gauge theory reflection of the fact that 
$SU(N|N)$ is reducible but indecomposable. 
Although we can follow Bars suggestion \cite{bars},
modifying the algebra in the gauge field sector to remove this feature,
and thus also $\A^0$, we cannot do so without destroying the
Leibnitz property of the usual representation of the 
super Lie bracket and thus gauge invariance
in the superscalar sector. Nevertheless the end result is that there are 
equivalent representations: the free-$\A^0$ representation already 
described (and the one we choose to work in because it is more elegant) and
a *bracket representation in which the $\A^0$ is `projected out'.
As we have seen, an alternative way of making sense of the $\A^0$ integral
is to work instead with $U(N|N)$. In all cases the bottom line is that only
the information in the supertraceless {\sl and} traceless part of the
$SU(N|N)$ superalgebra is actually propagated by the gauge fields $\A_\mu$.

The superscalar, $\C$, introduced to break the theory along all and
only fermionic directions must be a representation of $U(N|N)$ since it
has to break along $\sig3$. At first sight we are able to impose a
gauge invariant linear constraint on the coefficient field $\C^\sigma$,
but we note that this leads either to inconsistency or results in
further constraints, which this time are non-linear.  As we note in
\sec{SUNN}, the $SU(N|N)$ invariance of the theory is built on the
cyclicity property of the supertrace for supermatrices. (We have taken
care to make this manifest throughout.) The introduction of superghosts
with opposite statistics breaks this cyclicity property. An elegant
solution is to introduce a separate ghost grading, recalling that it is
actually a matter of choice whether different fermionic flavours
commute or anticommute. This also allows the usual form of the BRST
transformations to be a symmetry and thus ensures that the usual
required properties of gauge fixing (gauge independence, transversality
of on-shell Green functions {\it etc.}) hold.

Another point of principle, dealt with in \sec{Unitarity}, is the
meaning of the wrong sign action that appears as a consequence of the
supertrace for both $A^2$ and $C^2$. This does not signal an
instability of the theory since both kinetic and interaction terms have
the wrong sign. Guided by invariance under the supergroup we show in
\sec{Unitarity} that the result is mathematically well defined, however
requiring an indefinite metric Hilbert space. Fortunately, in the
continuum limit $\Lambda\to\infty$, $C^2$ becomes infinitely heavy, and
$A^2$ decouples in the way described in \sec{Unitarity}. 

As a matter of fact in the most interesting case of $D=4$ dimensions,
the wrong sign in $A^2$ sector implies that its $\beta$ function is
that of a trivial, rather than asymptotically free, theory. In the
continuum limit in which the bare coupling $g$ is sent in the usual way
logarithmically to zero as $\Lambda\to\infty$ (in order to achieve a
finite interacting $A^1$ theory), the $A^2$ sector loses all
interactions and becomes a free theory \cite{erg2}. 

In \sec{Spontaneously}, the full spontaneously broken action is
introduced, regularised with polynomials of covariant higher
derivatives, ranks $r$ and $\rt$, for the $\A$ and $\C$ parts
respectively. The superghosts are regulated by a polynomial of higher
derivatives (not covariant) of rank $\rh$ introduced through the gauge
fixing function.

The proof that the result is ultraviolet finite starts in
\sec{Counting}.  Here we establish the necessary and sufficient
constraints required on the ranks of the polynomials, such that the
maximum number of Feynman diagrams are superficially finite simply by
power counting. Furthermore, we show by finding one-loop examples that
they are necessary even after taking into account cancellations
resulting from supersymmetry.  The constraints are given for the
dimensions of interest, in fact for all $D\ge1$, by \eq{finalcons} and
agree precisely with the relevant inequalities proved in the manifestly
gauge invariant but incomplete formulation of ref.\ \cite{erg2}.
 
In this way, we are left to consider only a set of One-loop Remainder
Diagrams which are not finite purely by power counting,
namely those formed from only $\C^2\!\A^j$, $\A^j$ and $\etab\A\eta$ 
vertices, with $D$ or less external $\A$ legs and no external $\C$ or 
(anti)ghost legs. In \sec{Supertrace} we establish that for one-particle
irreducible one-loop
diagrams all the contributions associated with the spontaneous symmetry
breaking are already finite by power counting. All the 
remaining contributions, equal to those in the symmetric phase,
appear as the product of two supertraces over the external fields.
This is shown by first demonstrating that for the symmetric phase,
all the tree contributions appear as a single supertrace, and then
showing that on closure into one-loop diagrams, the single supertrace
always splits into two supertraces. At first sight this pattern is violated
by gauge-sector (\ie gauge and ghost) 
propagators, which also introduce  ordinary trace
terms as a consequence of completeness relations over the associated
both-supertraceless-and-traceless generators. However, this
is just another symptom of the strange r\^ole of $\A^0$, and once
we take into account properly that gauge-sector fields interact only through
commutators, all the ordinary trace terms cancel out.
Since $\str\A=0$ and $\str\one=0$, we thus immediately find that 
for $E_\A\le3$ external $\A$s, the symmetric
phase contributions vanish,
and thus the full broken phase contributions are finite.
The fact that, consistent with the structure of the supergroup, only the
supertrace (and never the trace) appears in the final result, means
that in the large $N$ limit the symmetric phase has no quantum corrections
at all. In particular this means that all One-loop Remainder Diagrams
are finite in the large $N$ limit, and thus the full spontaneously
broken theory is finite, in any dimension $D$.

Returning to finite $N$,
in \sec{Ward} we tackle the remaining contributions not already shown to
be superficially finite. These are symmetric phase one-loop vertices
with $3<E_\A\le D$ external gauge field legs. The key here is to take
properly into account the gauge invariance of the theory. Thus gauge invariance
tells us that any longitudinal part of the
four-point vertex vanishes, since it
is given by a sum over three-point vertices.
By power-counting the remaining
fully transverse four-point vertex is finite in all 
dimensions $D<8$. Iterating to higher
point vertices we thus establish finiteness of all these remaining 
contributions, and thus also the full theory, in all dimensions $D<8$.
However this argument, based as it is on an exact implementation of
gauge invariance (as in ref.\ \cite{erg1,erg2}), is not rigorous
in the present context: we have to worry that new divergences may occur
in terms including ghosts. To check this, we develop in standard fashion,
the full Lee-Zinn-Justin identities and check that the required BRST sources
introduce no divergences. Apart from some unimportant finite corrections
the argument above may then be repeated, with the same conclusion: at
finite $N$, the full theory is ultraviolet finite in all dimensions $D<8$.

In \sec{Unitarity}, we turn to the other crucial requirement of
a regulator: that in the limit $\Lambda\to\infty$ we are indeed left
with the theory we set out to regulate, namely $SU(N)$ Yang-Mills
theory carried by $A^1$. Actually in this limit we find
$SU(N)\times SU(N)$, with the second $SU(N)$ being carried by $A^2$.
We have to show then that $A^1$ and $A^2$ decouple, so that we can simply 
ignore the
$A^2$ sector. In the large $N$ limit, no interaction is possible since
only single trace interactions are allowed. At finite $N$, we show by
a rather straightforward application of the Appelquist-Carazzone
theorem that 
decoupling takes place providing that the theory is renormalizable,
\ie providing the dimension $D\le4$. We also note that decoupling fails
at finite $N$ in $D>4$ dimensions. This completes the proof that our
spontaneously broken $SU(N|N)$ theory
acts as a regulator for $SU(N)$ Yang-Mills, at least
to all orders in perturbation theory, for any dimension $D\le4$,
and in the large $N$ limit for all dimensions $D$.

In \sec{Preregularisation} we discussed in detail
the issue of preregularisation. 
Since in common with other Pauli-Villars approaches, finiteness is
achieved only after subtracting separately divergent contributions,
in general the answers are well defined only after applying and removing a
`preregulator' \cite{pvp}. The obvious choice of preregulator is in
effect to use dimensional regularisation by keeping the dimension $D$
general, taking the limit to the actual spacetime dimension only at the
end of the calculation \cite{erg2}. We stressed that
the preregulator is not used to compute divergences (there are none) or
to renormalize the theory, therefore the usual issues of defining what
dimensional regularisation means non-perturbatively do not arise.
We also showed that for all the cases $D<4$ and/or $N=\infty$,
no preregularisation is needed, the structure of the supergroup being
enough to cast all quantum corrections in terms of absolutely convergent
momentum integrals. Only in the (however important) case when the 
dimension $D=4$, and $N$ is finite, do we need to use preregularisation. 

We now draw some further conclusions. 
Despite much effort \cite{thooftln,lnefforts}, 
four dimensional large $N$ Yang-Mills
theory has evaded solution. However we saw in \sec{Supertrace}, 
that in the large $N$ limit the symmetric phase of this $SU(N|N)$
theory has no quantum corrections at all. It is therefore trivially
exactly soluble. (This is of course equally true of the pure $SU(N|N)$ 
Yang-Mills theory, \ie without the superscalar sector.)
Surely, this ought to help understand the large $N$ 
limit for $SU(N)$ Yang-Mills, which as we have seen is recovered at 
energies much less than $\Lambda$ in the spontaneously broken theory?
Note however that the large $N$ limit of the spontaneously broken
theory is not the same as implementing spontaneous symmetry breaking 
after the large $N$ limit has been taken. The two procedures do not
commute, as can be confirmed in the appropriate large $N$ 
variables:\footnote{which in particular requires $\Lambda^{D-2}$ to be replaced
by $N\Lambda^{D-2}$ in \eq{unbCact}} whereas extra supertrace factors of strings 
of $n$ superscalars $\str(\C\cdots\C)$, formally count as order one and
thus are 
subleading in the $1/N$ expansion, replacing these by their expectation
values $\Lambda\sig3$ results for odd $n$, in a leading contribution $\sim N$.

Furthermore we can note that, trivially, the large $N$ limit of
symmetric phase of this $SU(N|N)$ theory (or the pure case without
superscalars) is finite without the introduction of covariant higher 
derivatives. This is not the case for the spontaneously broken theory
as follows from the large $N$ limit of two of the one-loop examples below
\eq{1looprels2} used to prove the necessity of inequalities \eq{finalcons}.
At finite $N$, we see from the  $\C^4$ one-loop example, that the symmetric
phase also needs covariant higher derivatives (or some other regularisation)
to be ultraviolet finite. Likewise,
from the analysis of the two-loop graphs and higher, we also expect that
at finite $N$, pure $SU(N|N)$ Yang-Mills needs further regularisation
\eg the
covariant higher derivatives, to make it ultraviolet finite.

Finally, let us note that pure $SU(N|N)$ Yang-Mills and 
the symmetric phase of the theory has in common with ref. \cite{erg2},
a duality under 
\bea
\hbar &\mapsto& -\hbar\nonumber\\
\A_\mu &\mapsto& \sigma_1\A_\mu\sigma_1,\nonumber\\
\C &\mapsto& \sigma_1\C\sigma_1
\eea
which exchanges the r\^ole of $A^1$ and $A^2$ (and similarly $C^1$
and $C^2$;  $\sigma_1$ is defined in \eq{sig1}, and of course the
last equation is ignored for pure $SU(N|N)$ Yang-Mills.) By the
usual changes of variables on the fields to bring $g$ outside the
action, the change of sign on the loop counting parameter $\hbar$
becomes $g^2\to-g^2$ \cite{erg2}. Unlike the version in ref.
\cite{erg2}, this theory-space symmetry is broken once $\C$ picks
up an expectation value.
However as in ref. \cite{erg2}, the duality is also broken by the
differing renormalization required for the $A^1$ and $A^2$ sector (as
noted earlier).

Having established that this regularisation framework really works, at
finite $N$, to all orders in perturbation theory, and presumably thus
also non-perturbatively, the stage is now set to generalise the
manifestly gauge invariant exact renormalization group methods of refs.
\cite{manerg,erg1,erg2}, to computations in higher
order perturbation theory (\eg to check consistency)
and non-perturbatively, allowing 
for the first time such continuum computations to be performed 
without gauge fixing \cite{us}.

\section*{Acknowledgments} We would like to thank L. Alvarez-Gaum\'e,  
J. Kalkkinen, J.I. Latorre, H. Osborn, G. Shore and K. Stelle and 
especially A.A. Slavnov for discussions, and acknowledge
financial support from the PPARC grant PPA/V/S/1998/00907 and the Royal
Society Short Term Visitor grant ref. RCM/ExAgr/hostacct (T.R.M.,
Yu.K.), the Russian Foundation for Basic Research (grant 00-02-17679)
(Yu.K.), PPARC SPG PPA/G/S/1998/00527 (T.R.M., S.A.) and a PPARC
studentship (J.F.T.).

\appendix

\section{Background: the covariant Landau gauge controversy}
\label{Landau}

Slavnov's original formulation of covariant Pauli-Villars 
regularisation \cite{pvo} caused some controversy 
\cite{pvp,pvr,pvc}, not the least because, as noted by Martin and Ruiz Ruiz 
\cite{pvc}, some conflicting responses \cite{pvp,pvr} were qualitative and 
unchecked by explicit calculations, but also by the fact that, as described in 
\sec{Introduction},  there were actually several subtle problems
which were not always sufficiently distinguished: 
``overlapping divergences'' at higher loops, the imposition of a 
``covariant Landau gauge'', and the necessity of preregularisation. It appears 
that this has sometimes left the
impression that there is something inherently wrong with the whole 
idea of gauge invariant Pauli-Villars regularisation. Indeed, Martin and
Ruiz Ruiz took great care to expose precisely where the problem of 
preregularisation lay, and took the trouble to compute the one-loop
$\beta$ function using dimensional preregularisation, obtaining the wrong
answer \cite{pvc}. These authors correctly identified unphysical
contributions from the PV sector as causing the problem. 

Less well known it seems, is that the restriction to covariant Landau
gauge caused these unphysical contributions \cite{pva}, not the
existence of the PV fields themselves. As we further stress below, it
does so by hiding a massless mode, which does not decouple as the
regulator scale $\Lambda$ is sent to infinity \cite{pva}, thus
violating a basic requirement of any regulator system (\cf the comments
in \sec{Recovery}).  At the same time we demonstrate explicitly that
there is no analogue of this problem in our $SU(N|N)$ regularisation
scheme.

Before reviewing this, we stress some important related points. 

Firstly, none of these problems remain as an issue either for Slavnov's
scheme \cite{pvs} or the one we set out here. In particular, the
restriction to covariant Landau gauge in Slavnov's scheme, proved not
to be necessary and thus this difficulty was readily resolved
\cite{pva,pvs}.  Our scheme has no unregulated overlapping divergences,
since we have proved its finiteness to all loop orders (and see
\sec{Introduction}). No preregularisation is required for the cases
$D<4$ and/or $N=\infty$, whilst dimensional preregularisation is a
consistent choice for the remaining case of finite $N$ and $D=4$ (\cf
\sec{Preregularisation}).

Secondly, when the one-loop $\beta$ function was recomputed in the
repaired scheme the right answer was obtained \cite{pva}. In what is
essentially the $SU(N|N)$ regularisation described in this paper, the
one-loop $\beta$ function has also already been computed, using
dimensional preregularisation, yielding the right answer \cite{erg2}.

\medskip
 
In Slavnov's original paper \cite{pvo}, his 
PV fields $q_\mu(x)=q^a_\mu\tau^a$, $\tau^a$ being the generators of $SU(N)$,
were restricted to be orthogonal to the gauge covariant
derivative:
\be
\label{covLgauge}
D_\mu\cdot q^\mu=0,
\ee
by inserting the functional
$\delta(D_\mu\cdot q^\mu)$ into the partition function.
($D_\mu=\partial_\mu-igA_\mu$, and $A_\mu$ is the $SU(N)$ gauge field.
This is the so-called ``covariant Landau gauge'' \cite{pva}.) 
Taking the limit of large mass $\Lambda$ for the PV fields $q^\mu$ we
should return to a partition function for just $A$ governed by its
covariant higher derivative regularised Yang-Mills action $S_{YM}$.
(Here we are ignoring the gauge fixing and ghost terms for $A$, and
other PV fields \cite{pvo}: what we have will be sufficient to display
the problem \cite{pva}.) Actually, transferring the constraint
\eq{covLgauge} to the action by introducing a Lagrange multiplier field
$b=b^a\tau^a$, we obtain:
\be
\label{SlavnovZ}
{\cal Z}=\int\!\!{\cal D}(A,b,q)\ {\rm e}^{-S_{YM}-S_{PV}}
\ee
where 
\be
\label{SlavnovPV}
S_{PV} = \tr\!\!\int\!\! d^{D}\!x\left( 
2i\Lambda bD_\mu q^\mu +\Lambda^2q^2\right).
\ee
(Of course the factor of $2\Lambda$ in the $b$-$q$ term is harmless. It
is included for convenience.) The PV fields $q^\mu$ appear bilinearly
with an $A$-dependent bilinear form, but we have recognized that in
their large mass limit we can ignore all but the mass term. 
We cannot ignore the $b$-$q$ term however, because
the resulting bilinear form acting on $(b,q^\mu)$ would then be
singular. This already shows that the contribution from
\eq{SlavnovPV} cannot be trivial. Performing the $q$ functional
integral, \eq{SlavnovZ} becomes
\be
{\cal Z}=\int\!\!{\cal D}(A,b)\ {\rm e}^{-S_{YM}-S_{PV}},
\ee
where now
\be
S_{PV} = \tr\!\!\int\!\! d^{D}\!x \left(D_\mu\cdot b\right)^2
\ee
reveals the hidden massless mode. 

In refs. \cite{pva,pvs}, it was noted that the PV fields continue to
regulate if \eq{covLgauge} is replaced by a ``covariant $\alpha$
gauge'', \ie instead of using $b$ and the $b$-$q$ term in
\eq{SlavnovPV}, one adds a term ${1\over\alpha}(D_\mu\cdot q^\mu)^2$ to
the PV action.  Clearly this can be ignored relative to $\Lambda^2 q^2$
in the large $\Lambda$ limit: it does not result in a
singular bilinear form and the PV fields do fully decouple.  We see
that it is indeed the imposition of covariant Landau gauge that prevents the PV
fields from properly decoupling in their infinite mass limit, leaving
behind unphysical massless modes \cite{pva}.

Finally, returning to our own system of regularisation, it is easy to
confirm that in the infinite $\Lambda$ mass limit our PV fields $B$,
$C$ and $D$, and the off-diagonal ghosts $\psi$ and $\phi$, all gain
infinite masses (\cf \eq{brkCact}, \eq{gaugeact} and \eq{ghostact}
respectively). As we showed in \sec{Alternatives}, $\A^0$ does not
appear in the action. This leaves only the massless Yang-Mills fields
$A^1$, $A^2$, and their respective ghosts. We see that there are no
analogues of the covariant Landau gauge (or covariant $\alpha$ gauge for
that matter) and in particular no analogue of  $b$, the singular
$b$-$q$ terms and the unphysical contributions they caused.



\end{document}